\documentclass[prd,twocolumn,floatfix,superscriptaddress,showpacs,letterpaper,altaffilletter,nofootinbib]{revtex4}
\usepackage{color}
\usepackage{times}
\usepackage{graphicx}
\usepackage{fancyhdr}
\usepackage{float}
\usepackage{ulem}
\makeatletter
\makeatletter
\def\@fnsymbol#1{\ifcase#1\or * \or  $+$ \or  \$ \or \#  \or \dag \or \ddag \or
$\mathsection$ \or $ \mathparagraph$ \or $\|$  \or \textordfeminine \or \textbullet   
\or ** \or $++$ \or  \$\$ \or \#\#  \or \dag\dag \or \ddag\ddag \or
$\mathsection\mathsection$ \or $ \mathparagraph\mathparagraph$ \or $\|\|$  \or 
\textordfeminine\textordfeminine \or \textbullet \textbullet \or *** \or $+++$ 
\or  \$\$\$ \or \#\#  \or \dag\dag \or \ddag\ddag \or
$\mathsection \mathsection\mathsection$ \or $ \mathparagraph 
\mathparagraph\mathparagraph$ \or $\|\|\|$  \or 
\textordfeminine\textordfeminine\textordfeminine \or 
\textbullet\textbullet\textbullet \or \else \@ctrerr\fi}
\makeatother

\newcommand\checkme[1]{\textcolor{black}{{#1}}}

\newcommand\chisqthresh{\ensuremath{\xi^\ast}}

\newcommand\rundurationhours{1415}
\newcommand\LOneScidatahours{536}		
\newcommand\HOneScidatahours{1044}		
\newcommand\HTwoScidatahours{822}		
\newcommand\HOneHTwoOnlyScidatahours{385}	
\newcommand\TripleObservationHours{242}		
\newcommand\LOneHOneObservationHours{99}	
\newcommand\LOneHTwoObservationHours{32}	
\newcommand\TotalObservationHours{373}		

\newcommand\totaltimehours{\mbox{$339$}}

\newcommand\rhomaxsqr{\mbox{$89.1$}}

\newcommand\rpermweg{\mbox{$44.4$}}
\newcommand\ratepermweg{\mbox{$47$}}
\newcommand\nmweg{\mbox{$1.34$}}

\newcommand\plusminus[2]{\mbox{$+$#1$/$$-$#2}}

\def\thercsid{\relax}

\renewcommand{\today}{\number\day\space\ifcase\month\or
  January\or February\or March\or April\or May\or June\or
  July\or August\or September\or October\or November\or December\fi
  \space\number\year}

\begin{document}

\title{Search for gravitational waves from galactic and extra--galactic binary neutron stars\}
}

\date[\relax]{ RCS \thercsid; compiled \today }

\begin{abstract}\quad
We use 373 hours ($\approx$ 15 days) of data from the second science
run of the LIGO gravitational-wave detectors to search for signals
from binary neutron star coalescences within a maximum distance of
about 1.5~Mpc, a volume of space which includes the
Andromeda Galaxy and other galaxies of the Local Group of galaxies.
This analysis requires a signal to be found in data from detectors at
the two LIGO sites, according to a set of coincidence criteria.  The
background (accidental coincidence rate) is determined from the data
and is used to judge the significance of event candidates.
No inspiral gravitational wave events were identified in our search.
Using a population model which includes the Local Group, we establish
an upper limit of less than \ratepermweg ~inspiral events per
year per Milky Way equivalent galaxy with 90\% confidence for 
non-spinning binary
neutron star systems with component masses between 1~and 3~$M_\odot$.
\end{abstract}

\newcommand*{\AG}{Albert-Einstein-Institut, Max-Planck-Institut f\"ur Gravitationsphysik, D-14476 Golm, Germany}
\affiliation{\AG}
\newcommand*{\AH}{Albert-Einstein-Institut, Max-Planck-Institut f\"ur Gravitationsphysik, D-30167 Hannover, Germany}
\affiliation{\AH}
\newcommand*{\AN}{Australian National University, Canberra, 0200, Australia}
\affiliation{\AN}
\newcommand*{\CH}{California Institute of Technology, Pasadena, CA  91125, USA}
\affiliation{\CH}
\newcommand*{\DO}{California State University Dominguez Hills, Carson, CA  90747, USA}
\affiliation{\DO}
\newcommand*{\CA}{Caltech-CaRT, Pasadena, CA  91125, USA}
\affiliation{\CA}
\newcommand*{\CU}{Cardiff University, Cardiff, CF2 3YB, United Kingdom}
\affiliation{\CU}
\newcommand*{\CL}{Carleton College, Northfield, MN  55057, USA}
\affiliation{\CL}
\newcommand*{\FN}{Fermi National Accelerator Laboratory, Batavia, IL  60510, USA}
\affiliation{\FN}
\newcommand*{\HC}{Hobart and William Smith Colleges, Geneva, NY  14456, USA}
\affiliation{\HC}
\newcommand*{\IU}{Inter-University Centre for Astronomy  and Astrophysics, Pune - 411007, India}
\affiliation{\IU}
\newcommand*{\CT}{LIGO - California Institute of Technology, Pasadena, CA  91125, USA}
\affiliation{\CT}
\newcommand*{\LM}{LIGO - Massachusetts Institute of Technology, Cambridge, MA 02139, USA}
\affiliation{\LM}
\newcommand*{\LO}{LIGO Hanford Observatory, Richland, WA  99352, USA}
\affiliation{\LO}
\newcommand*{\LV}{LIGO Livingston Observatory, Livingston, LA  70754, USA}
\affiliation{\LV}
\newcommand*{\LU}{Louisiana State University, Baton Rouge, LA  70803, USA}
\affiliation{\LU}
\newcommand*{\LE}{Louisiana Tech University, Ruston, LA  71272, USA}
\affiliation{\LE}
\newcommand*{\LL}{Loyola University, New Orleans, LA 70118, USA}
\affiliation{\LL}
\newcommand*{\MP}{Max Planck Institut f\"ur Quantenoptik, D-85748, Garching, Germany}
\affiliation{\MP}
\newcommand*{\MS}{Moscow State University, Moscow, 119992, Russia}
\affiliation{\MS}
\newcommand*{\ND}{NASA/Goddard Space Flight Center, Greenbelt, MD  20771, USA}
\affiliation{\ND}
\newcommand*{\NA}{National Astronomical Observatory of Japan, Tokyo  181-8588, Japan}
\affiliation{\NA}
\newcommand*{\NO}{Northwestern University, Evanston, IL  60208, USA}
\affiliation{\NO}
\newcommand*{\SC}{Salish Kootenai College, Pablo, MT  59855, USA}
\affiliation{\SC}
\newcommand*{\SE}{Southeastern Louisiana University, Hammond, LA  70402, USA}
\affiliation{\SE}
\newcommand*{\SA}{Stanford University, Stanford, CA  94305, USA}
\affiliation{\SA}
\newcommand*{\SR}{Syracuse University, Syracuse, NY  13244, USA}
\affiliation{\SR}
\newcommand*{\PU}{The Pennsylvania State University, University Park, PA  16802, USA}
\affiliation{\PU}
\newcommand*{\TC}{The University of Texas at Brownsville and Texas Southmost College, Brownsville, TX  78520, USA}
\affiliation{\TC}
\newcommand*{\TR}{Trinity University, San Antonio, TX  78212, USA}
\affiliation{\TR}
\newcommand*{\HU}{Universit{\"a}t Hannover, D-30167 Hannover, Germany}
\affiliation{\HU}
\newcommand*{\BB}{Universitat de les Illes Balears, E-07122 Palma de Mallorca, Spain}
\affiliation{\BB}
\newcommand*{\BR}{University of Birmingham, Birmingham, B15 2TT, United Kingdom}
\affiliation{\BR}
\newcommand*{\FA}{University of Florida, Gainesville, FL  32611, USA}
\affiliation{\FA}
\newcommand*{\GU}{University of Glasgow, Glasgow, G12 8QQ, United Kingdom}
\affiliation{\GU}
\newcommand*{\MU}{University of Michigan, Ann Arbor, MI  48109, USA}
\affiliation{\MU}
\newcommand*{\OU}{University of Oregon, Eugene, OR  97403, USA}
\affiliation{\OU}
\newcommand*{\RO}{University of Rochester, Rochester, NY  14627, USA}
\affiliation{\RO}
\newcommand*{\UW}{University of Wisconsin-Milwaukee, Milwaukee, WI  53201, USA}
\affiliation{\UW}
\newcommand*{\WU}{Washington State University, Pullman, WA 99164, USA}
\affiliation{\WU}

\author{B.~Abbott}    \affiliation{\CT}
\author{R.~Abbott}    \affiliation{\LV}
\author{R.~Adhikari}    \affiliation{\LM}
\author{A.~Ageev}    \affiliation{\MS}  \affiliation{\SR}
\author{B.~Allen}    \affiliation{\UW}
\author{R.~Amin}    \affiliation{\FA}
\author{S.~B.~Anderson}    \affiliation{\CT}
\author{W.~G.~Anderson}    \affiliation{\TC}
\author{M.~Araya}    \affiliation{\CT}
\author{H.~Armandula}    \affiliation{\CT}
% ...

 \collaboration{The LIGO Scientific Collaboration, http://www.ligo.org}
 \noaffiliation

\maketitle

\section{Introduction}

The search for gravitational waves has entered a new era with the
scientific operation of kilometer scale laser interferometers.  These
L-shaped instruments are sensitive to minute changes in the relative
lengths of their orthogonal arms that would be produced by
gravitational waves \cite{Saulson:1994}.  The Laser Interferometer
Gravitational-wave Observatory (LIGO)
\cite{Barish:1999,LIGOS1instpaper} consists of three
Fabry-Perot-Michelson interferometers: two interferometers are housed
at the site in Hanford, WA; a single interferometer is housed in
Livingston, Louisiana.
In 2003, all three instruments simultaneously collected data under
stable operating conditions during two science runs.  Even though the
instruments were not yet performing at their design sensitivity, the
data represent the best broad-band sensitivity to gravitational waves
that has been achieved to date. 

In this paper, we report the methods and results of a search for
gravitational waves from binary neutron star systems, using
data from the science run conducted early in 2003.  These waves are
expected to be emitted at frequencies detectable by LIGO during the
final few seconds of inspiral as the binary orbit decays due to the
loss of energy in gravitational radiation~\cite{Cutler:1992tc}.  A
previous search~\cite{LIGOS1iul}, using data from the first LIGO
science run, reported an upper limit on the rate of coalescences
within our Galaxy and the Magellanic clouds.  This paper uses
an analysis pipeline which is optimized for detection by using data only
from times when interferometers were operating properly at both LIGO
sites.  By demanding that a gravitational wave be seen at both sites, we
strongly suppress the rate of
background events from non-astrophysical disturbances.  Moreover, 
this approach allows us to judge the significance of any apparent event
candidate in the context of the background distribution,
which is also determined from the data. 

The search described here has essentially perfect efficiency 
for detecting binary neutron star
inspirals within the Milky Way and the Magellanic Clouds 
(as measured by Monte Carlo simulations), and
could detect some inspirals as far away as the Andromeda
and Triangulum galaxies (M31 and M33).  The rate of coalescences in
these galaxies, based on the population of known binary neutron star
systems~\cite{Kalogera:2004tn}, is expected to be very low, so that a
detection by the present search would be highly surprising.  In fact,
no coincident event candidates were observed in excess of the measured
background.  The data are therefore used to place an improved direct
observational upper limit on the rate of binary neutron star
coalescence events in the Universe.

\section{Data Sample}
\label{s:datasample}

The LIGO Hanford Observatory (LHO) in Washington state has two
independent detectors sharing a common vacuum envelope, one with 4~km
long arms (H1) and one with 2~km long arms (H2).  The LIGO Livingston
Observatory (LLO) in Louisiana has one detector with 4~km long arms
(L1).  All three detectors operated during the second LIGO science
run, referred to as S2, which spanned 59 days from February 14 to
April 14, 2003.  During operation, feedback to the mirror positions
and to the laser frequency keeps the optical cavities near resonance,
so that interference in the light from the two arms recombining at the
beam splitter is strongly dependent on the difference between the
lengths of the two arms.  A photodiode at the antisymmetric port of
the detector senses this light, and a digitized signal is recorded at
a sampling rate of 16384~Hz.  This channel can then be searched for a
gravitational wave signal. More details on the detectors' instrumental
configuration and performance can be found in \cite{LIGOS1instpaper}
and \cite{LIGOS2grb}.

\begin{figure}
\begin{center}
\hspace*{-0.2in}\includegraphics[width=\linewidth]{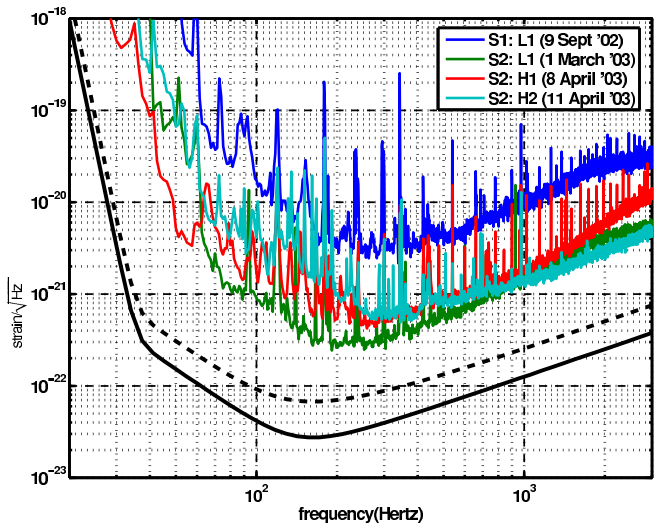}
\end{center}
\caption{\label{f:S1S2curves}
Typical sensitivities, expressed as amplitude spectral densities of
detector noise converted to equivalent gravitational-wave strain, for
the best detector during the S1 run (the Livingston detector) and for
all three detectors during the S2 run.
The solid lower line is the design sensitivity for the LIGO 4-km
detectors;  the dashed line is the design sensitivity for the LIGO
2-km detector at Hanford.  }
\end{figure}

While the detailed noise spectrum of a detector affects different
gravitational wave searches in different ways, we can summarize the
sensitivity of a detector for low-mass inspiral signals in terms of
the range for an archetypal source.  Specifically, the range is the
distance at which an optimally oriented and located binary
system\footnote{An optimally oriented and located binary system would
be located at the detector's zenith with its orbital plane
perpendicular to the line of sight between the detector and the
source.} with the mass of each component equal to $1.4 M_\odot$ would
yield an amplitude signal-to-noise ratio (SNR) of 8 when extracted
from the data using optimal filtering.
During the first LIGO science run (referred to as S1),
the L1 detector had the greatest range, typically $0.18$~Mpc.  During
the S2 run, all three detectors were substantially more
sensitive than this, with ranges of $2.0$, $0.9$,
and $0.6$~Mpc for L1, H1, and H2 averaged over all times during the
run.  Typical amplitude
spectral densities of detector noise are shown in
Fig.~\ref{f:S1S2curves}.

The amount of science data with good performance and
stable operating conditions was limited by environmental factors
(especially high ground motion at LLO and strong winds at LHO),
occasional equipment failures, and periodic special investigations.
Over the \rundurationhours\ hour duration of the S2 run, the total amount of science
data obtained was \LOneScidatahours\ hours for L1, 
\HOneScidatahours\ hours for H1, and 
\HTwoScidatahours\ hours for H2.

The analysis presented here uses data collected while the
LLO detector was
operating at the same time as one or both of the LHO detectors.
Science data during which both H1 and H2 were operating but L1 was
not, amounting to 
\HOneHTwoOnlyScidatahours\ hours, was not used in this analysis
because of concerns about possible environmentally-induced
correlations between the data streams of these two co-located
detectors; this data set, as well as data collected while only one of
the LIGO detectors was in science mode, will be used in a separate
analysis together with data from the TAMA300
detector~\cite{TAMA:2001}, which conducted ``Data Taking~8''
concurrently with the LIGO S2 run.

\begin{figure}
\begin{center}
\hspace*{-0.2in}\includegraphics[width=0.7\linewidth]{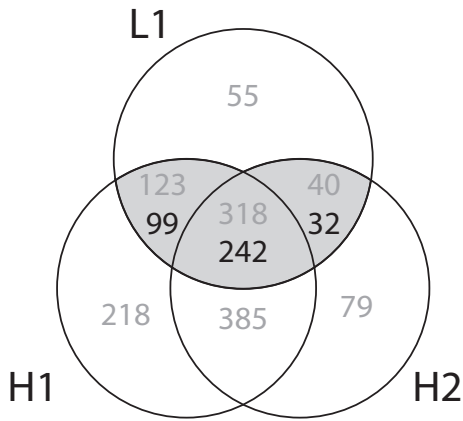}
\end{center}
\caption{\label{f:S2times}
The number of hours that each
detector combination was operational during the S2 run.  The upper
number gives the amount of time the specific instruments were
coincidentally operational.  The lower number gives the total
time that was searched for inspiral triggers.  The
shaded region corresponds to the data used in this search.}
\end{figure}

Of all the data, approximately 9\% (uniformly sampled within the run
so as to be representative of the whole data set) was used as
{\it playground} data for tuning parameters and thresholds in
the analysis pipeline, and for use in identifying vetoes that were
effective in eliminating spurious events.
This playground data set was excluded from the
gravitational wave inspiral upper limit calculation because event
selection and pipeline tuning, as described in Secs.~\ref{s:triggers},
\ref{s:vetoes} and \ref{s:pipeline}, introduces statistical bias which
cannot be accounted for.   The playground data set was
searched for inspiral signals, however, so a potential detection during these times
was not excluded.
After applying data quality cuts (as detailed in
Sec.~\ref{s:vetoes}) and accounting for short time intervals which
could not be searched for inspiral signals by the filtering algorithm
(described in Sec.~\ref{s:triggers}), the observation time
consisted of 
\TripleObservationHours\ hours of triple-detector data, plus
\LOneHOneObservationHours\ hours of L1-H1 and 
\LOneHTwoObservationHours\ hours of L1-H2 data, for
a total observation time of 
\TotalObservationHours\ hours. For the upper limit
result, the non-playground observation time was 
\totaltimehours\ hours.
A summary of the amount of single, double and triple detector times is
provided in Fig.~\ref{f:S2times}.

As with earlier analyses of LIGO data \cite{LIGOS1iul}, the output of
the antisymmetric port of the detector was calibrated to obtain a
measure of the relative strain $s=\Delta L/L$ of the detector arms,
where $\Delta L=L_x-L_y$ is the difference in length between one
arm (the $x$ arm) and the other (the $y$ arm), and $L$ is the average
arm length.  Reference calibration functions, tracing out the
frequency-dependent response of the detectors, were measured (by moving
the end mirrors of the detector with a known displacement) before and
after the science run, and once during the science run;  all three
measurements gave consistent results. 
The changing optical gain of the calibration was
monitored continuously during the run by applying sinusoidal motions
with fixed frequency to the end mirrors.  This continuous monitoring,
averaged once a minute, allowed for small corrections to the
calibration due to loss of light power in the arms, which can be
caused by drifting optical alignment.

\section{Target Sources}
\label{s:population}

Binary neutron star systems in the Milky Way, known from radio pulsar
observations~\cite{Stairs:2004}, provide indirect evidence for the
existence of gravitational waves~\cite{weisberg:2002}. Based on
current astrophysical understanding, the spatial distribution of
binary neutron stars is expected to follow that of star formation in
the Universe. A measure of star formation is the blue
luminosity of galaxies appropriately corrected for dust extinction and
reddening.  Therefore we model the spatial distribution of double
neutron stars according to the corrected blue-light distribution of
nearby galaxies~\cite{Phinney:1991ei}.
While the masses of neutron stars in the few known binary systems are
all near $1.4 M_\odot$, population synthesis simulations suggest that
some systems will have component masses as low as $\sim 1 M_\odot$ and
as high as the theoretical maximum neutron star mass of $\sim 3
M_\odot$~\cite{Belczynski:2002}.  Thus, we search for inspiral signals
from binary systems with component masses in this range.  Note that
the higher-mass systems radiate more energy in
gravitational waves and can thus be detected at a greater distance
at a given SNR.  For component masses below $1 M_\odot$,  a search is
reported in Ref.~\cite{LIGOs2macho}.

When the LIGO detectors reach their design sensitivities, they will be
capable of detecting inspiral signals from thousands of galaxies,
reaching beyond the Virgo Cluster for systems with optimal location
and orientation.  At that sensitivity, the rate of detectable binary
neutron star coalescences could be as high as $0.7$ per year, though
it is more likely to be an order of magnitude smaller
\cite{Kalogera:2004tn}.
For this analysis, our target population includes the Milky Way
and all significant galaxies within a distance of 3~Mpc,
which is roughly the maximum distance for which a $3$--$3 M_\odot$
inspiral could be detected in coincidence by the L1 and H1
interferometers with a SNR of 6 in H1.
This population includes the Local Group of galaxies, whose total
blue-luminosity is
dominated by the Andromeda Galaxy (M31), as well as some galaxies from
neighboring groups.  We cannot hope to detect {\em all}
inspirals within this volume, because most systems in the population
have lower masses and because the received signal amplitude is
reduced, on average, depending on the orientation and location of the
source relative to the detector.

Table~\ref{t:galaxies} gives the parameters we use for the galaxies in
the target population out to 1.5~Mpc, i.e., the maximum distance for which we
had a non-zero detection efficiency in our simulations. 
The coordinates and distances of the galaxies in
Table~\ref{t:galaxies} are taken from a catalog by
Mateo~\cite{Mateo:1998}, when available; this catalog is favored
because distances quoted are from individual, focused studies of each
of the nearby galaxies he includes. The rest of the distances, with
only 100 kpc accuracy, are taken from the Tully Nearby Galaxies
catalog~\cite{Tully:1994}. Data for blue luminosities are derived from
the apparent blue magnitudes (corrected for reddening) quoted in
Ref.~\cite{RC3:1991}, and the distances shown in
Table~\ref{t:galaxies}. We measured the efficiency of our search
using Monte Carlo simulations, where the sources in the target
population had a mass distribution as described in
Ref.~\cite{Belczynski:2002}, following the same guidelines as in the
population models used in Ref.~\cite{Nutzman:2004}. We used simulations
with a population of neutron stars from galaxies up to 3~Mpc away,
over-extending our target population, although we did not detect any
simulated injections from sources farther than 1.5~Mpc away.

\begin{figure}
\includegraphics[width=\linewidth]{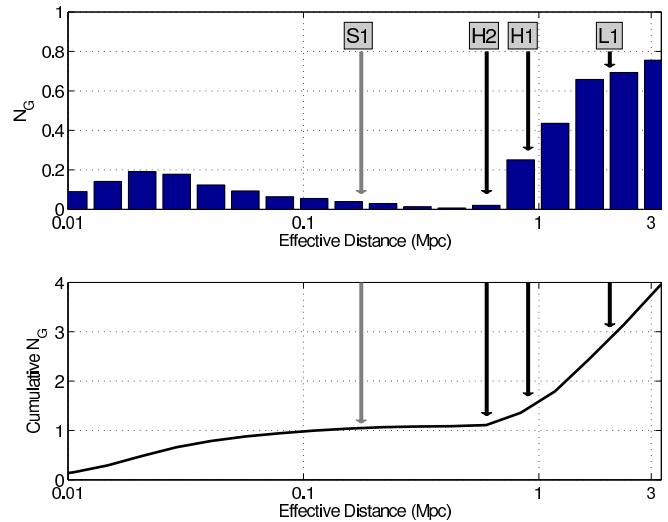}
\caption{\label{f:population} The upper panel shows the histogram of
the number of Milky Way Equivalent Galaxies ($N_G$) as a function of
{\it effective} distance.  Labeled arrows indicate the average range
of the best detector in the first science run (S1), as well as the
average range of each detector during the second science run. The
cumulative total within a given effective distance is shown in the bottom panel.}
\end{figure}

\begin{table*}
\caption{\label{t:galaxies} The galaxies in our population within
1.5~Mpc.  The next to last column indicates the number of injections
detected in coincidence (with signal-to-noise-ratio defined in Eq.~\ref{e:combined-snr} satisfying $\rho^2>89$), over the
number of injections performed in our simulations. The last column
indicates the cumulative number of equivalent Milky Way galaxies contributed by
systems within the corresponding distance, calculated from the search efficiency
and the blue light luminosity.}
\vspace*{0.1in}
\newdimen\digitwidth\setbox0=\hbox{$0$}
\digitwidth=\wd0\catcode`?=\active
\def?{\kern\digitwidth}
\begin{ruledtabular}
\begin{tabular}{lccccccc}
  	& 	Right Ascension 	& 	Declination 	& 	Distance 	& 	Blue light luminosity 	& 	Detected/Injected 	& 	Cumulative 	\\ 
Name 	& 	(Hour:Min) 	& 	(Deg:Min) 	& 	(kpc) 	& 	 relative to Milky Way 	& 	with $\rho^2>89$ 	& 	$N_G$ 	\\ \hline 
Milky Way 	& 	--- 	& 	--- 	& 	--- 	& 	1 	& 	686/686 	& 	1 	\\ 
LMC 	& 	+05:23.6 	& 	-69:45 	& 	49 	& 	0.128 	& 	57/57 	& 	1.128 	\\ 
SMC 	& 	+00:52.7 	& 	-72:50 	& 	58 	& 	0.037 	& 	8/8 	& 	1.165 	\\ 
NGC6822 & 	+19:44.9 	& 	-14:49 	& 	490 	& 	0.01 	& 	0/4 	& 	1.165 	\\ 
NGC185 	& 	+00:39.0 	& 	+48:20 	& 	620 	& 	0.007 	& 	1/3 	& 	1.1673 	\\ 
M110 	& 	+00:40.3 	& 	+41:41 	& 	815 	& 	0.018 	& 	5/100 	& 	1.1682 	\\ 
M31 	& 	+00:42.7 	& 	+41:16	& 	770 	& 	2.421 	& 	108/1791 & 	1.3142 	\\ 
M32 	& 	+00:42.7 	& 	+40:52	& 	805 	& 	0.019 	& 	9/129 	& 	1.3155 	\\ 
IC10 	& 	+00:20.4 	& 	+59:18	& 	825 	& 	0.031 	& 	1/10 	& 	1.3186 	\\ 
M33 	& 	+01:33.9 	& 	+30:39	& 	840 	& 	0.319 	& 	9/219 	& 	1.3318 	\\ 
NGC300 	& 	+00:54.9 	& 	-37:41	& 	1200 	& 	0.052 	& 	1/40 	& 	1.3331 	\\ 
M81 	& 	+09:55.6 	& 	+69:04	& 	1400 	& 	0.196 	& 	1/147 	& 	1.3344 	\\ 
NGC55 	& 	+00:14.9 	& 	-39:11	& 	1480 	& 	0.175 	& 	2/122 	& 	1.3373 	\\ 
\end{tabular}
\end{ruledtabular}
\end{table*}

Within the LIGO frequency band, the gravitational waveform produced 
by binary neutron star systems is well described by the
restricted second-post-Newtonian approximation~\cite{Apostolatos:1995, Droz:1997, Droz:1998}. The spins of the
neutron stars are not expected to significantly affect the orbital
motion or the waveform~\cite{Apostolatos:1995}.  Tidal 
coupling and other finite-body effects dependent on the equation of state
also are not expected to significantly affect the waveform
in the LIGO band~\cite{Bildsten:1992my}. 
The waveform received at Earth is therefore
parameterized by the masses of the companions, the
distance to the binary, the inclination of the system to the plane of
the sky, and by the initial orbital phase when the waveform enters the LIGO
band.  The waveforms consist of two polarizations, the plus ($h_+$) and
cross ($h_\times$) polarizations, which describe the two orthogonal tidal
distortions produced by the waves.  These polarization basis states are
defined with respect to the orientation of the binary orbit relative to the
line of sight.  An interferometric detector is
sensitive to a particular linear combination of these two polarizations;
this is described by two response functions $F_+$ and $F_\times$ so that the
expected gravitational-wave signal is
\begin{equation}
  h(t) = F_+ h_+(t) + F_\times h_\times(t).
\end{equation}
The response functions depend on the location of the detector on the Earth and
on the orientation of the detector's arms.
The interferometers at LHO and LLO are aligned as closely as possible, but
the curvature of the Earth causes a slight difference in their
antenna patterns.

\section{Filtering and Trigger Generation} 
\label{s:triggers} 
We generated event {\em triggers} by filtering the data $s(t)$ from
each detector with matched filters designed to detect the expected
signals.  For any given binary neutron star mass pair, $\{m_1,m_2\}$,
we constructed the expected frequency-domain inspiral waveform
{template}, $\tilde{h}(f)$, using a stationary phase approximation to
the restricted second post Newtonian waveform~\cite{Blanchet:1996pi}.\footnote{The 
stationary-phase approximation to the Fourier transform of inspiral 
template waveforms was shown to be sufficiently accurate for gravitational-wave 
detection in Ref.~\cite{Droz:1999qx}.}
Here, the tilde indicates the Fourier transform of a time
series $h(t)$ according to the convention
\begin{equation} 
  \tilde{h}(f) = \int_{-\infty}^\infty h(t) e^{-2\pi ift}dt.  
\end{equation} 
The matched filter output is then the complex time series
\begin{equation}
\label{e:xfilter}
  z(t) = x(t) + iy(t)= 4\int_{0}^{\infty}
    \frac{\tilde{h}^\ast(f)\tilde{s}(f)}{S(f)}\,e^{2\pi ift} df
\end{equation}
where $x(t)$
is the (real) matched filter output for the inspiral waveform
with a zero orbital phase, and $y(t)$ is the (real)
matched filter output for the inspiral waveform with a
$\pi/2$ orbital phase.  The quantity $S(f)$ is the one-sided
strain noise power spectral density, estimated from the data.  The matched
filter variance is given by
\begin{equation}\label{e:sigmasq} 
  \sigma^2 = 4\int_0^\infty \frac{|\tilde{h}(f)|^2}{S(f)} df 
\end{equation} 
which depends on the template's amplitude normalization.  The
amplitude signal-to-noise ratio (SNR) is then
\begin{equation}\label{e:rho}
  \rho=|z|/\sigma \, .
\end{equation}

The templates were normalized to a binary neutron star inspiral
at an {\em effective distance} of 1~Mpc, where the effective distance
of a waveform is the distance for which a binary neutron star system
would produce the waveform if it were optimally oriented.  Thus, 
\begin{equation}\label{effdist}
  D_{\mathrm{eff}} = \frac{\sigma}{\rho}
\end{equation} 
is an estimate of the effective distance, in Mpc, of a putative signal
that produces SNR $\rho$.  

Since each binary neutron star mass pair $\{m_1,m_2\}$ would produce a
slightly different waveform, we constructed a {\em bank} of templates
with different mass pairs such that, for any actual mass pair with
$1\,M_\odot\le m_2\le m_1\le 3\,M_\odot$, the loss of SNR due to the
mismatch of the true waveform from that of the best fitting waveform
in the bank is less than $3\%$~\cite{Owen:1995tm,Owen:1998dk}.

Although a threshold on the matched filter output $\rho$ would be the
optimal detection criterion for an accurately-known inspiral waveform
in the case of stationary, Gaussian noise, the character of the data
used in this analysis is known to be neither stationary nor Gaussian.
Indeed, many classes of transient instrumental artifacts have been
categorized, some of which produce copious numbers of spurious,
large SNR events.  In order to reduce the number of spurious event
triggers, we adopted a now-standard $\chi^2$
requirement~\cite{Allen:2004}: The matched filter template is divided
into $p$ frequency bands, which are chosen so that each band would
contribute a fraction $1/p$ of the total SNR if a true signal (and no
detector noise) were present. In our analysis, we used $p=15$, as
explained in Sec.~\ref{ss:tuning}. We then construct a
chi-squared statistic comparing the magnitude and phase of SNR
accumulated in each band to the expected amount.  For a true signal in
Gaussian noise, the resulting statistic is $\chi^2$
distributed with $2p-2$ degrees of freedom (since the SNR is
constructed out of the two matched filters $x$ and $y$ which are both
measured).  Instrumental artifacts
tend to produce very large $\chi^2$ values and can be rejected
by requiring $\chi^2$ to be less than some reasonable threshold.
Due to the discreteness of the template bank, however, a real signal
will generally not match precisely the nearest template in the bank.
Consequently, $\chi^2$ has a non-central chi-squared distribution with
non-centrality parameter $\lambda \le (p-1) \rho^2 \mu^2$, where
$\mu$ is the fractional loss of SNR due to mismatch between
template and signal~\cite{Allen:2004}.  For sufficiently small
$\mu$ and moderate values of $\rho$, this effect is not
important, but when $\rho$ is large, it must be taken into account.
This is done by applying a threshold with the following
parametrization:
\begin{equation}\label{chisqthresh} 
  \chi^2 \le {(p + \delta^2\rho^2)} \chisqthresh \, .
\end{equation}
The threshold multiplier $\chisqthresh$, the number of bins $p$ 
and the value of $\delta$ were
determined by tuning on the playground data, as will be described in
Sec.~\ref{ss:tuning}.

In this analysis, the SNR $\rho(t)$ was computed for each template in
the bank.  Whenever $\rho(t)$ exceeded a threshold $\rho^{\ast}$, the
value of $\chi^{2}$ was computed for that time.  If $\chi^{2}$ was
below the threshold in Eq.~(\ref{chisqthresh}), then the local
maximum of $\rho(t)$ was recorded as a {\em trigger}. Each trigger is
represented by a vector of values: the masses which define the
template, the maximum value of $\rho$ and corresponding value of
$\chi^{2}$, the inferred coalescence time, the effective distance
$D_{\mathrm{eff}}$, and the coalescence phase $\phi_{0} =
\tan^{-1}(y/x)$.

\section{Data Quality Checks and Vetoes}
\label{s:vetoes}
In practice, the performance of the matched filtering algorithm
described above is influenced by non-stationary optical alignment,
servo control settings, and environmental conditions.  We used two
strategies to avoid problematic data in this search.  One was to
evaluate {\it data quality} over relatively long time intervals, using
several different tests. Time intervals identified as being suspect or
demonstrably bad were skipped when filtering the data.  The other
method was to look for signatures in environmental monitoring channels
and auxiliary interferometer channels which would indicate an external
disturbance or instrumental glitch (a large transient fluctuation),
allowing us to {\it veto} any triggers recorded at that time.

Several data quality tests were applied {\it a priori}, leading us to
omit data when 
calibration
information was missing or unreliable, servo control settings were not
at their nominal values, or there were input/output controller timing
problems.  Additional tests were performed to characterize the noise
level in the interferometer in various frequency bands and to check
for problems with the photodiodes and associated electronics.  The
playground data set was used to judge the relevance of these
additional tests, and two data quality tests were found to correlate
with inspiral triggers found in the playground.  One of these
pertained only to the H1 interferometer; there were occasional, abnormally
high noise levels in the H1 antisymmetric port channel
that were apparent when
this signal was averaged over a minute.  Data was rejected only if
this excessive noise was present for at least 3 consecutive minutes.
The other data quality test used to reject data pertained to
saturation of the photodiode at the antisymmetric port. These
photodiode saturation events correlated with a small, but significant,
number of L1 inspiral triggers. We required the absence of photodiode
saturation in the data from all three detectors.

A gravitational wave would be most evident in the signal obtained at
the antisymmetric port. We examined many other auxiliary
interferometer channels, which monitor the light in the interferometer
at points other than the antisymmetric port, to
look for correlations between glitches found in the readouts of these
channels and inspiral event triggers found in the playground data.  
These auxiliary
channels are sensitive to certain instrumental artifacts which may
also affect the antisymmetric port, and therefore may provide highly
effective veto conditions.  Although these channels are
expected to have little or no sensitivity to a gravitational wave, we
considered the possibility that an actual astrophysical signal
could produce the observed glitches in the auxiliary channel due to
some physical or electronic coupling.  This possibility was tested by
means of {\em hardware} injections, in which a simulated inspiral
signal is injected into the data by physically moving one of
the end mirrors of the interferometer. These hardware injections were
used for validation of the analysis pipeline, as described in
Sec.~\ref{ss:validation}. Unlike the software injections that are
used to measure the pipeline efficiency (described in
Sec.~\ref{ss:tuning}), these hardware injections allow us to
establish a limit on the effect that a true signal would have on the
auxiliary channels.  Only those channels that were unaffected by the
hardware injections were considered safe for use as potential veto
channels.

We used an analysis program, {\it glitchMon}, \cite{glitchMon} to
identify large amplitude transient signals in auxiliary channels.
Numerous channels were examined by glitchMon, which generates a list
of times when glitches occurred, identified by a filtered time series
crossing a chosen threshold.
A veto condition based on a given list of glitchMon triggers was
defined by choosing a fixed time window around each glitch and
rejecting any inspiral event trigger with a coalescence time within
the window.

For each veto condition considered, we evaluated the 
veto efficiency (percentage
of inspiral events eliminated), use percentage (percentage of veto
triggers which veto at least one inspiral event), and dead-time
(percentage of science-data time eliminated by the veto). Once a
channel was identified, some tuning was done of the filters used, and
the thresholds and time windows chosen, to optimize the efficiency,
especially for high SNR candidates, without an excessive deadtime. The
parameter tuning was done only with the inspiral triggers found in the
playground.

No efficient candidate veto channels were identified for H1 and H2;
there were some candidates for L1.  Non-stationary noise in
the low-frequency part of the sensitivity range used for the
inspiral search appeared to be the dominant cause for glitch events in
the data. The original frequency range for the binary neutron star
inspiral search extended from $50$~Hz to $2048$~Hz.  It was discovered
that many of the L1 inspiral triggers appeared to be the result of
non-stationary noise with frequency content around $70$~Hz.  An
important auxiliary channel, L1:LSC-POB\_I, proportional to the
length fluctuations of the power recycling cavity, was found to have
highly variable noise at $70$~Hz.  
As a consequence, it was decided that the low-frequency cutoff of the
binary neutron star inspiral search should be increased from 50 Hz to
$100$~Hz.  This subsequently reduced the number of inspiral
triggers. An inspection of artificial signals injected into the data
revealed a very small loss of efficiency for binary neutron star
inspiral signal detection resulting from the increase in the
low-frequency cutoff.

Even after raising the low-frequency cutoff, the L1:LSC-POB\_I channel
was found to be an effective veto when filtered appropriately. 
Due to the characteristics of the inspiral templates' response to
large glitches, we decided to veto with a very wide time window,
$-4$~s to $+8$~s, around the time of each L1:LSC-POB\_I trigger.
With this choice, 12~\% of the BNS inspiral triggers with $\rho>8$ in
the playground were vetoed, as well as 5 out of the 9 triggers found
in the playground with $\rho>10$. The use percentage of the veto
triggers was 18~\% for $\rho>8$ and 0.7~\% for $\rho>10$,
whereas values of 3~\% and $<0.1$~\% (respectively) would be expected
from random coincidences if the veto triggers had no real correlation
with the inspiral triggers.
The total L1 dead-time using this veto condition in the playground was
2.7~\%.  The performance in the full data set was consistent
with that found in the playground: the final observation time
(including playground) was reduced by this veto from 385
hours to 373 hours. A more extensive discussion of LIGO's S2
binary inspiral veto study can be found in \cite{vetoGWDAW03}.

\section{Search for Coincident Event Candidates} \label{s:pipeline}
\subsection{Analysis Pipeline}
\label{ss:analysispipeline}

The detection of a gravitational-wave inspiral signal in the S2 data
would (at the least) require triggers in L1 and one or more of the
Hanford instruments with consistent arrival times (separated 
by the light travel time between the detectors) and waveform
parameters.  Requiring temporal coincidence between the two
observatories greatly reduces the background rate due to spurious
triggers, thus allowing an increased confidence for detection candidates. 
When detectors at both observatories are operating
simultaneously, we may obtain an estimate of the rate of background
triggers by time-shifting the Hanford triggers with respect to the
Livingston 
triggers and applying the same coincidence requirements to the
time-shifted triggers,  as described in Sec.~\ref{s:background}. In this
way, we can measure the rate of accidental coincidences in our
search.

\begin{figure}
\begin{center}
\hspace*{-0.2in}\includegraphics[width=0.9\linewidth]{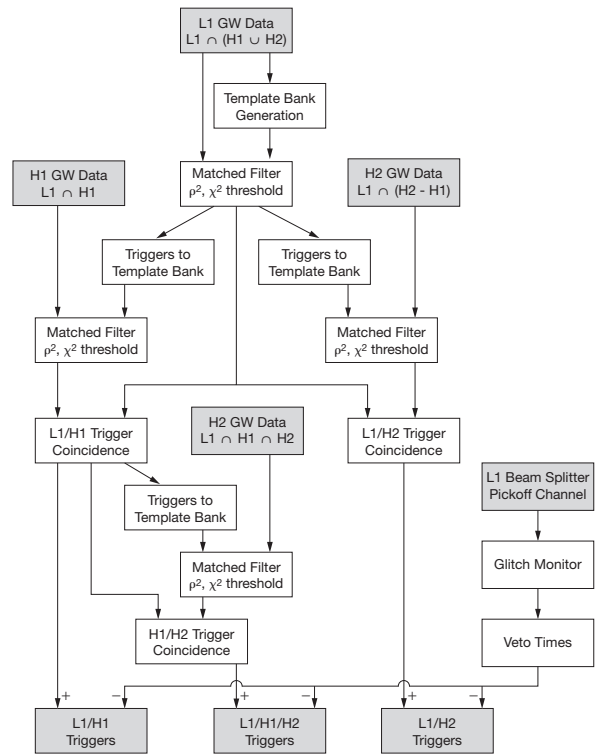}
\end{center}
\caption{\label{f:pipeline}
The inspiral analysis pipeline used to determine the reported upper
limit. $\mathrm{L1} \cap (\mathrm{H1} \cup \mathrm{H2})$ indicates times when
the L1 interferometer was operating in coincidence with one or both of the
Hanford interferometers. $\mathrm{L1} \cap \mathrm{H1}$ indicates times when
the L1 interferometer was operating in coincidence with the H1 interferometer.
$\mathrm{L1} \cap (\mathrm{H2} - \mathrm{H1})$ indicates times when the L1
interferometer was operating in coincidence with only the H2 interferometer.
The outputs of the search pipeline are triggers that belong to one of the
two double-detector coincident data sets or to the triple-detector data set.}
\end{figure}

During the S2 run, the three LIGO detectors had substantially
different sensitivities. The sensitivity of the L1 detector was
greater than that of either Hanford detector throughout the run. Since
the orientations of the LIGO interferometers are similar, we expect
that signals of astrophysical origin detected in the Hanford
interferometers generally are detectable in the L1 interferometer.
Using this as a guiding principle, we have constructed a { triggered
search} pipeline, summarized in Fig.~\ref{f:pipeline}. We search for
inspiral triggers in the most sensitive interferometer (L1), and only
when a trigger is found in this interferometer do we search for a
coincident trigger in the less sensitive interferometers. This
approach reduces significantly the computational power necessary to
perform the search, without compromising the detection efficiency of
the pipeline. 

Times when each interferometer was in stable operation are called
science segments.  The science segments from each interferometer are
used to construct three data sets corresponding to: (1) times when all
three interferometers were operating, (2) times when only the L1 and
H1 interferometers were operating; and (3) times when only the L1 and
H2 interferometers were operating. The pipeline produces a list of
{coincident triggers} for each of these three data sets as described
below.

The science segments were analyzed in chunks of 2048 seconds
using the \textsc{findchirp} implementation of matched filtering for
inspiral signals in the LIGO Algorithm library~\cite{LALS2BNS}.  In this
code, the data
set for each 2048 second chunk is first down-sampled from 16384~Hz to
4096~Hz.  It is subsequently high-pass filtered and a low frequency
cutoff of 100~Hz imposed.  The calibrated instrumental response for
the chunk is calculated using the average of the calibrations
(measured every minute) over the duration of the chunk.

\begin{figure}
\begin{center}
\hspace*{-0.2in}\includegraphics[width=0.9\linewidth]{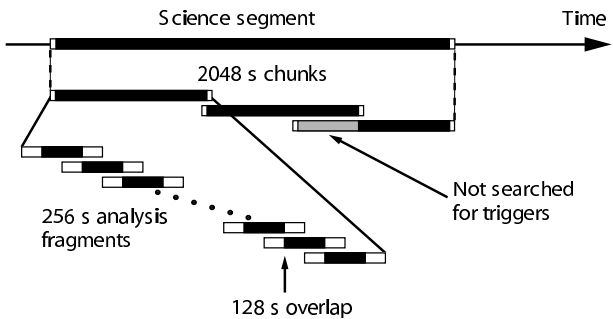}
\end{center}
\caption{\label{f:chunks}
The algorithm used to divide science segments into data analysis
fragments.  Science segments are divided into $2048$~s chunks
overlapped by $128$~s. (Science segments shorter than $2048$~s are
ignored.) An additional chunk with a larger overlap is added to cover
any remaining data at the end of a science segment.  Each chunk is
divided into $15$ analysis fragments of length $256$~s for
filtering. The first and last $64$~s of each analysis fragment are
ignored, so the segments overlap by $128$~s.  Areas shaded black are
searched for triggers by the search pipeline. The gray area in the
last chunk of the science segment is not searched for triggers as this
time is covered by the preceding chunk, although these data points are
used is used in estimating the noise power spectral density for the
final chunk.}
\end{figure}

Triggers are not searched for
within the first and last $64$~s of a given chunk, so subsequent
chunks are overlapped by $128$~s to ensure that all of the data in a
continuous science segment (except for the first and last 64
seconds) are searched for triggers.  Any science segments shorter than
$2048$~s are ignored. If a science segment cannot be exactly divided
into overlapping chunks (as is usually the case) the remainder of the
data set is covered by a special $2048$-s chunk which overlaps with the
previous chunk as much as necessary to allow it to reach the end of
the segment.  For this final chunk, a parameter is set to restrict the
inspiral search to the time interval not covered by any previous chunk,
as shown in Fig~\ref{f:chunks}.

Each chunk is further split into 15 analysis fragments of length 256
seconds overlapped by 128 seconds.  The power
spectrum $S(f)$ for the 2048 seconds of data is estimated by taking
the median of the power spectra of the 15 segments.  (We use the
median and not the mean to avoid biased estimates due to large
outliers, produced by non-stationary data.) The average
calibration is applied to the data in each analysis fragment, and the
matched filter output in Eq.~(\ref{e:xfilter}) is computed for each
template in the template bank.

In order to avoid end-effects when applying the matched filter, the
frequency-weighting factor, nominally $1/S(f)$, is altered so that its
inverse Fourier transform has a maximum duration of $\pm 16$ seconds.
The output of the matched filter near the beginning and end of each
segment is corrupted by end-effects due to the finite duration of the
power spectrum weighting and also the inspiral template.  By ignoring
the filter output within 64~s of the beginning and end of each segment,
we ensure that only uncorrupted filter output is searched for inspiral
triggers.  This necessitates the overlapping of segments and chunks as
described above.

The power spectral density (PSD) of the noise in the Livingston
detector is estimated independently for each L1 chunk that is
coincident with operation of a Hanford detector [denoted $\mathrm{L1}
\cap (\mathrm{H1} \cup \mathrm{H2})$].  The PSD is used to lay out a
template bank for filtering that chunk, according to the parameters
for mass ranges and minimal match~\cite{Owen:1995tm, Owen:1998dk}. The
data from the L1 interferometer for the chunk are then filtered, using
that bank, with a signal-to-noise threshold $\rho_{\mathrm{L}}^\ast$
and $\chi^2$ veto threshold $\chisqthresh_{\mathrm{L}}$ to produce a
list of triggers as described in Sec.~\ref{s:triggers}.  For each
chunk in the Hanford interferometers, a {\em triggered bank} is
created consisting of every template which produced at least one
trigger in L1 during the time of the Hanford chunk.  This is used to
filter the data from the Hanford interferometers with signal-to-noise
and $\chi^2$ thresholds specific to the interferometer, giving a total
of six thresholds that may be tuned.  For times when only the H2
interferometer is operating in coincidence with L1 [denoted
$\mathrm{L1} \cap (\mathrm{H2} - \mathrm{H1})$] the triggered bank is
used to filter the H2 chunks that overlap with L1 data; these triggers
are used to test for L1/H2 coincidence.  All H1 data that overlaps
with L1 data (denoted $\mathrm{L1} \cap \mathrm{H1}$) are filtered
using the triggered bank for that chunk. For H1 triggers produced
during times when all three interferometers were operating, a second
triggered bank is produced for each H2 chunk consisting of every
template which produced at least one trigger found in coincidence in
L1 and H1 during the time of the H2 chunk. The H2 chunk is
filtered with this bank.  Any H2 triggers found with this bank are
tested for triple coincidence with the L1 and H1 triggers.  The
remaining triggers from H1, when H2 is not available, are used to search
for L1/H1 coincident triggers.

\begin{figure}
\begin{center}
\hspace*{-0.2in}\includegraphics[width=0.9\linewidth]{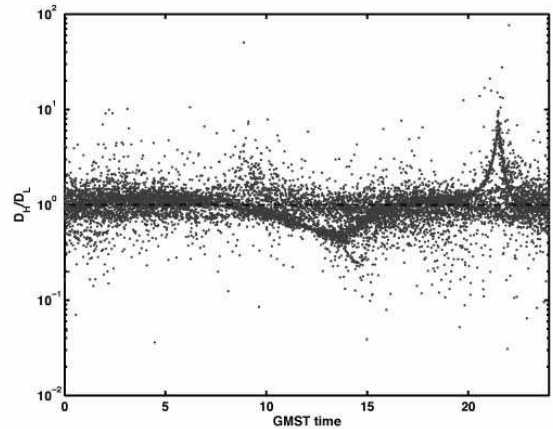}
\end{center}
\caption{\label{f:eff_dist_ratio_hist}
Ratio of effective distance at the Hanford and Livingston
Observatories for the injections from sources in
Table.~\ref{t:galaxies}, versus Greenwich mean Sidereal Time (GMST).
The sharp feature near 21.5 hrs is due to M31 (Andromeda) passing
through a sky position at the Hanford node, producing much larger
effective distances at LHO than at LLO. The softer feature near 14 hrs
is due to M33 (Triangulum Galaxy) passing through a similar
sensitivity node for LLO.  About 15\% of injected signals have a
$50\%$ difference or larger in the effective distance between the
sites due to the slight mis-alignment of the detectors.  }
\end{figure}

For a trigger to be considered coincident between two interferometers,
the following conditions must be fulfilled: (1) Triggers must be
observed in both interferometers within a temporal coincidence window
that allows for the error in measurement of the time of the trigger,
$\delta t$. If the detectors are not co-located, this parameter is
increased by the light travel time between the observatories ($10$~ms
for traveling 3000~km at the speed of light). (2) We then ensure that
the triggers have consistent waveform parameters by demanding that the
two mass parameters for the template are identical to within an error
of $\delta m$.  (3) For H1 and H2, we may impose an amplitude cut on
the triggers given by
\begin{equation}
\label{eq:eff_dist_test}
\frac{\left|D_\mathrm{1} - D_\mathrm{2}\right|}{D_\mathrm{1}}
< \frac{\epsilon}{\rho_\mathrm{2}} + \kappa,
\end{equation}
where $D_1$ ($D_2$) is the effective distance of the trigger in the
first (second) detector and $\rho_{2}$ is the signal-to-noise ratio of
the trigger in the second detector. The parameters $\epsilon$ and
$\kappa$ are tunable.   As shown
in Fig.~\ref{f:eff_dist_ratio_hist}, the non-perfect alignment of LLO
and LHO (due to their different latitudes) can occasionally cause
large variations in the detected signal amplitudes for astrophysical
signals.  In order to disable the amplitude cut when
comparing triggers from LLO and LHO, we set $\kappa = 1000$. 

If the detectors are at the same site, we ask if the maximum distance
to which H2 can see at the signal-to-noise threshold
$\rho_\mathrm{H2}^\ast$ is greater than the distance of the H1
trigger, allowing for errors in the measurement of the trigger
distance. If this is the case, we demand time, mass and effective
distance coincidence.  If the distance to which H2 can see overlaps the
error in measured distance of the H1 trigger, we search for a trigger
in H2, but always keep the H1 trigger even if no coincident trigger is
found. If the minimum of the error in measured distance of the H1
trigger is greater than the maximum distance to which H2 can detect a
trigger we keep the H1 trigger without searching for coincidence.

If coincident triggers are found in H1 and H2, we can get an improved
estimate of the amplitude of the signal arriving at the Hanford site
by coherently combining the filter outputs from the two gravitational
wave channels, 
\begin{equation}
\rho_H = \sqrt{ \frac{|z_{H1} + z_{H2}|^2}{\sigma_{H1}^2 +
     \sigma_{H2}^2} } \; .
\label{e:rhoH}
\end{equation}
The more sensitive interferometer receives more weight in this
combination, as can be seen from Eq.~(\ref{e:xfilter}), in which
the noise enters in the denominator.
If a trigger is found in only one of the Hanford interferometers, then
$\rho_H$ is simply taken to be the value of $\rho$ from that interferometer.

The final step of the search is to apply the POB\_I veto,
described in Sec.~\ref{s:vetoes}, to eliminate certain L1 triggers
which arose from instrumental glitches.  The individual
signal-to-noise ratios are used to construct a multi-detector
statistic as described in Sec.~\ref{s:results} for any surviving
triggers.

The surviving coincident triggers are clustered in a way that
identifies the best parameters to associate with a possible inspiral
signal in the data.  The clustering is needed since large
astrophysical signals and instrumental noise bursts can produce many
triggers with coalescence times within a few seconds of each other.
We chose the trigger with the largest SNR from each cluster; triggers
separated by more than 4 seconds were considered unique. Alternative
clustering methods are discussed in Sec.~\ref{s:background} in an
effort to understand the accidental likelihood of a small number of
candidates that were observed in the final sample.

To perform the search on the full data set, a directed acyclic graph (DAG) was
constructed to describe the work flow, and execution of the pipeline tasks
was managed by Condor~\cite{beowulfbook-condor} on the UWM and LIGO Beowulf
clusters. The software to perform all steps of the analysis and construct the
DAG is available in the package \textsc{lalapps}\cite{LALS2BNS}.
     
\subsection{Parameter Tuning}
\label{ss:tuning}

The entire analysis pipeline was studied first using the playground
data set to tune the values of the various parameters. The
goal of tuning was to maximize the efficiency of the
pipeline to detection of gravitational waves from binary inspirals
without producing an excessive rate of spurious candidate events.  The
detection efficiency was determined by Monte Carlo
simulations in which we injected simulated inspiral signals from our
model population into the data. The efficiency is the ratio
of the number of signals detected to the number injected, as described in
Sec.~\ref{sss:single_ifo_tuning}.  In the absence of a detection, a
pipeline with a high efficiency and low false alarm rate allows us to
set a better upper limit, but it should be noted that our primary
motivation is to enable reliable detection of gravitational waves.
By evaluating the efficiency using Monte Carlo injection of
signals from the hypothetical population of binary neutron stars into
the data,  we account for systematic effects caused by our vetoes and 
other aspect of our pipeline.

There are two sets of parameters that we can tune in the pipeline: (1)
the single interferometer parameters which are used in the matched filter and
$\chi^2$ veto to generate inspiral triggers in each interferometer, and (2) the
coincidence parameters used to determine if triggers from two interferometers
are coincident. The single interferometer parameters include the
signal-to-noise threshold $\rho^\ast$, the number of frequency
sub-bands $p$ in the $\chi^2$ statistic, the coefficient $\delta$ on
the SNR dependence of the $\chi^2$ cut, and the $\chi^2$ cut threshold
$\chisqthresh$. These are tuned
on a per-interferometer basis, although some of the values chosen are common
to two or three detectors.  The coincidence parameters are the time
coincidence window for triggers, $\delta t$, the mass parameter coincidence
window, $\delta m$, and the effective distance cut parameters $\epsilon$ and
$\kappa$ described in Eq.~(\ref{eq:eff_dist_test}).  Due to the nature of the
triggered search pipeline, parameter tuning was carried out in two stages. We
first tuned the single interferometer parameters for the primary detector (L1).
We then used the triggered template banks (generated from the L1 triggers)
to explore the single
interferometer parameters for the less sensitive Hanford detectors. Finally,
the parameters of the coincidence test were tuned.

\subsubsection{Single Interferometer Tuning}
\label{sss:single_ifo_tuning}

The number of bins $p$ used in the $\chi^2$ test was set to $p=15$
(compared with $p=8$ used in our S1 search~\cite{LIGOS1iul}), in order
to better differentiate spurious triggers from actual (or injected)
signals, while still having at least several cycles of the waveform in
each bin.

The $\delta$ parameter in Eq.~(\ref{chisqthresh}) is expected to be no
less than 0.03 for our choice of maximum 3\% SNR loss in the L1
template bank. A
dedicated investigation using software injections into L1
showed that the $\chi^2$ test rejected 50\% of signals from the Milky
Way (with effective distances closer than 200 kpc) when using 
$\delta=0.03$; using $\delta=0.1$ recovered all such
injections. The search efficiency for weaker injections, at effective
distances larger than 900 kpc, did not depend on $\delta$.

The signal to noise threshold $\rho^\ast=6$ was used in all three
instruments.  This choice was motivated by the observation that a
signal,  with certain orbital orientations and sky positions,  can 
have a smaller effective distance in the
less sensitive detector (see next Sec~\ref{sss:coinc_tuning} and
Fig.~\ref{f:eff_dist_ratio_hist}).  This choice of threshold
was computationally possible due to our efficient pipeline and code; 
it also provided better statistics for background estimation
(Sec~\ref{s:background}).

Once the L1 search parameters had been tuned, the resulting triggered
template banks were used as input to tune the H1 and H2 $\chi^2$ threshold
parameters $\chisqthresh$ in the coincidence search.  The final value 
was selected so that the triggered search suffered no loss of efficiency
due to this single parameter.  Table~\ref{t:chisq_tune} shows how 
the parameter $\chisqthresh$ was tuned, first for L1 and then for
H1. The values chosen for all parameters are shown in Table~\ref{t:ifo_params}.

\begin{table}
\caption{\label{t:chisq_tune}
The effect of lowering the $\chi^2$ threshold, $\chisqthresh$, for the
single interferometer L1, and for the combination of L1 and H1 in the
pipeline.  The efficiency of the L1 search remains constant as the
threshold is lowered to $5.0$, however as the threshold is lowered in
H1, the efficiency of the triggered search drops. Further testing
indicated that a threshold of $12.5$ in H1 was acceptable without a
loss of efficiency.  }
\begin{tabular}{l|l|l}
Value of $\chi^2$ threshold, $\chisqthresh_{L1}$ & L1 Efficiency & Pipeline Efficiency \\
\hline 
20.0 & 0.350 & 0.270 \\
15.0 & 0.350 & 0.270 \\
10.0 & 0.350 & 0.255 \\ 
5.0 & 0.350 & 0.212 \\
\end{tabular}
\end{table}

\subsubsection{Coincidence Parameter Tuning}
\label{sss:coinc_tuning}

After the single interferometer parameters had been selected, the
coincidence parameters were tuned.  As described in
Sec.~\ref{ss:validation}, the coalescence time of an inspiral signal
can be measured to within $\le1$~ms. The gravitational waves' travel
time between observatories is $10$~ms, so $\delta t$ was chosen to be
$1$~ms for LHO-LHO coincidence and $11$~ms for LHO-LLO
coincidence. The mass coincidence parameter was initially chosen to be
$\delta m = 0.03$, however testing showed that this could be set to
$\delta m = 0.0$ ({\it i.e.}\ requiring the triggers in each
interferometer to be found with the exact same template) without loss
of efficiency.

After tuning the time and mass parameters, we tuned the effective
distance parameters $\kappa$ and $\epsilon$. Initial estimates of
$\epsilon = 2$ and $\kappa = 0.2$ were used for testing, however we
noticed that many injections were missed when testing for LLO-LHO
distance consistency.  This is due to the slight detector misalignment
between the two sites from Earth's curvature, which causes the ratio
of effective distances, as measured at the two observatories, to be
large for a significant fraction of our target population, as shown in
Fig.~\ref{f:eff_dist_ratio_hist}. Consequently, we disabled the
effective distance consistency requirement for triggers generated at
different observatories.  A study of simulated events injected into H1
and H2 suggested values of $\epsilon_{HH} = 2$ and $\kappa_{HH} = 0.5$
to be suitable. Note that, as described above, we demand that an L1/H1
trigger pass the H1/H2 coincidence test if the effective distance of
the trigger in H1 is within the maximum range of the H2 detector at
threshold.

\begin{table}
\caption{\label{t:ifo_params}
A complete list of the parameters that were selected at the various
stages of the pipeline. The procedures used to select these parameter values are
outlined in the text.  }
\begin{tabular}{clr}
Parameter & Pipeline node & value \\
\hline 
$\rho^\ast$ & GW Data (all) & 6.0 \\
$p$ & GW Data (all) & 15 \\
$\delta$ & GW Data (all) & 0.1 \\
$\chisqthresh_L$ & L1 GW Data & 5.0 \\
$\chisqthresh_H$ & H1/H2 GW Data & 12.5 \\
$\delta t_{HH}$ & H1/H2 Trigger Coincidence & 0.001 s \\
$\delta t_{HL}$ & L1/H1 and L1/H2 Trigger Coincidence & 0.011 s \\
$\delta m$ & Trigger Coincidence (all) & 0.0 \\
$\kappa_{HH}$ & H1/H2 Trigger Coincidence & 0.5 \\
$\epsilon_{HH}$ & H1/H2 Trigger Coincidence & 2 \\
$\kappa_{HL}$ & L1/H1 and L1/H2 Trigger Coincidence  &  1000
\end{tabular}
\end{table}

\subsection{Validation}
\label{ss:validation}

Hardware signal injections were used as a test of our data analysis
pipeline.  These injections allow us to study issues of instrumental
timing and calibration, as well as to verify that the injected signals
were indeed identified as triggers in our analysis pipeline.  At
intervals throughout S2, a predetermined set of inspiral and burst
signals were injected into the instruments using the mirror actuators.

We examined six sets of hardware injections spread throughout the S2
science run.  Each set had six $1.4\textrm{-}1.4 M_\odot$ inspiral
signals at effective distances spaced logarithmically between 500 kpc
and 15 kpc, and four $1.0\textrm{-}1.0 M_\odot$ inspirals at distances
from 500 kpc to 62 kpc.  Each strain waveform was calculated using the
second order post-Newtonian expansion for an optimally oriented
inspiral, and appropriately scaled for an inspiral at the desired
effective distance.  The strain was then converted into an injection
signal into the $y$-arm of the interferometer, with the appropriate
calibration to produce the desired differential strain.

Of these injections, five sets were injected simultaneously into all
three instruments and the sixth was injected into L1 and H1 only.
After the coincidence stage of the pipeline, 59 of the 60 injections
produced a trigger in the appropriate template ($1.4$--$1.4 M_\odot$
or $1.0$--$1.0 M_\odot$).  (The single missed injection had a $\chi^2$
value in L1 which was slightly above threshold.) This is an excellent
test of every stage of our pipeline.  L1 successfully produced
triggers corresponding to the injections, which were then used to make
triggered banks against which the Hanford instruments were analyzed.
Both H1 and H2 produced triggers which were coincident with those in
L1.  Furthermore, the amplitude cut between H1 and H2 was effective.
For the louder injections, the effective distance was consistent
between the two instruments.  In the case of more distant injections
which did not produce triggers in H2, the coincidence stage of the
pipeline allowed the L1--H1 coincident triggers to be kept even
though no event was found in H2.

The pipeline was also run on an additional set of hardware injections
at distances between 5 Mpc and 150 kpc. Of these injections, the most
distant which produced coincident triggers were at 1.25 Mpc for the
$1.4-1.4 M_{\odot}$ injection and 620 kpc for the $1.0-1.0 M_{\odot}$
injection.  This is consistent with the sensitivity of H1 during the
injection time, and with the efficiency measured in our software
injections (Table~\ref{t:galaxies}). 

The trigger times associated with the hardware injections all agreed
with expectations to within 1 msec.  

Hardware injections provide a powerful test of instrumental calibration.
Any errors in the calibration of the instrument affect the measured
effective distance to the inspiral.  Uncertainties in distance
measurements will contribute to errors in our detection efficiency.
Additionally, they become important when requiring consistency of
effective distance in triggers from the two Hanford instruments.  To
test the accuracy of the effective distance measurements, we used
the loudest injections --- namely the $1.4-1.4 M_\odot$ injections
with distances less than 100 kpc --- so that 
systematic errors in distance measurement would dominate over noise.  The
effective distance of all such injections was accurate to within $20\%$.
We found that for L1, the distance to injections was systematically
underestimated by $5\%$, with a $4\%$ standard deviation.  For H1, there
was a systematic underestimation of $2\%$ with a standard deviation of
$3\%$.  For H2, we had a $2\%$ overestimation and a standard
deviation of $5\%$.  These errors are consistent with uncertainties in
the calibration described in Sec.~\ref{s:errors}.  Further details of
the S2 hardware injections are available in \cite{S2Hardware:2004}.

A Markov chain Monte Carlo routine (MCMC) \cite{MCMC:2004} was also used
to examine the injected signals.  This provides a method of estimating
the parameters of an injected signal.  As an example, for a $1.4-1.4
M_\odot$ injection at 125 kpc, the MCMC routine generated masses of
$1.4003 M_\odot$ and $1.3991~M_\odot$.  The width of the 95~\%
confidence interval for each mass was $0.012~M_\odot$.

\section{Background estimation}
\label{s:background}

An event candidate which survives all cuts in our analysis pipeline,
including coincidence, can arise either from a real gravitational-wave
signal or from noise bursts which contaminate our data streams.  We
refer to the latter class of event candidates as background.  These
background events are caused by many different environmental and
instrumental processes.  Under the assumption that such processes are 
uncorrelated between the detectors at Hanford and Livingston, we
estimated the rate for background events due to accidental coincidences 
by applying artificial time shifts $\Delta t$
to the triggers coming from the Livingston detector.  These time-shift
triggers were then fed into subsequent steps of the pipeline.  For a
given time-shift, the triggers that survived to the end of the pipeline
represent a single trial output from our search (if no coincident
gravitational-wave signals were present).

\begin{figure}
\begin{center}
\includegraphics[width=\linewidth]{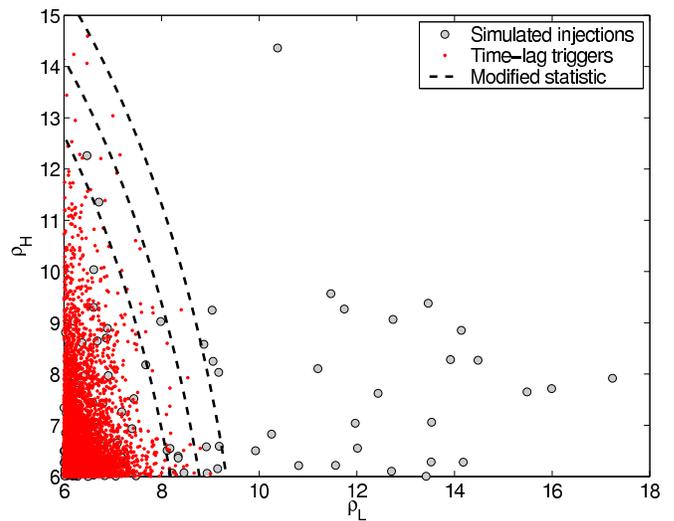}
\caption{\label{f:backgroundstat}
The signal to noise at Livingston $\rho_{\mathrm{L}}$ plotted against 
the signal to noise at Hanford $\rho_{\mathrm{H}}$ for triggers from
40 time-shifts.  In Gaussian noise,  one expects circular
false alarm contours centered on the origin indicating that the sum of
the squares of the signal to noise ratios would be a useful combined
statistic.  The observed distribution of time-shifted triggers motivated the
modified statistic presented in Eq.~(\protect\ref{e:combined-snr}).
}
\end{center}
\end{figure}

A total of 40 time-shifts were analyzed to estimate the
background:  $\Delta t = \pm 5$, $\pm 10$, $\pm 15$, $\pm 27$,
$\pm 37$, $\pm 47$, $\pm 57$, $\pm 67$, $\pm 77$, $\pm 87$, $\pm 97$,
$\pm 107$, $\pm 117$, $\pm 127$, $\pm 137$, $\pm 147$, $\pm 157$, $\pm
167$, $\pm 177$, $\pm 197$ seconds.  To avoid correlations, we used
time-shifts longer than the duration of the longest template waveform
($\sim 4 \textrm{ seconds}$).  We did not time-shift the triggers from
the Hanford detectors relative to one another since real correlations
may arise from environmental disturbances.  The resulting distribution
of time-shift triggers in the $(\rho_{\mathrm{H}},\rho_{\mathrm{L}})$
plane was used to determine a joint signal to noise $\rho$.  Here
$\rho_L$ is given by Eq. (\ref{e:rho}) and $\rho_H$ is given by Eq.
(\ref{e:rhoH}).  For Gaussian noise fluctuations, with the single
interferometer triggers' SNR maximized over the polarization phase and
the orbital inclination angle, one expects circular false alarm
contours centered on the origin~\cite{Pai:2000zt}, suggesting that the
sum of the squares of the signal to noise ratios would be a useful
combined statistic.  The time-shifts revealed the need for a modified
statistic; we settled on contours of (roughly) constant false-alarm
probability to use in assigning a signal-to-noise to coincident
triggers. As shown in Fig.~\ref{f:backgroundstat}, the combined SNR
\begin{equation}
\rho = \sqrt{ \rho^2_{\mathrm{L}} +  \rho_{\mathrm{H}}^2 / 4 }
\label{e:combined-snr}
\end{equation}
yielded approximate constant-density contours for the distribution of
background events in the plane.  

\begin{figure}
\begin{center}
\includegraphics[width=\linewidth]{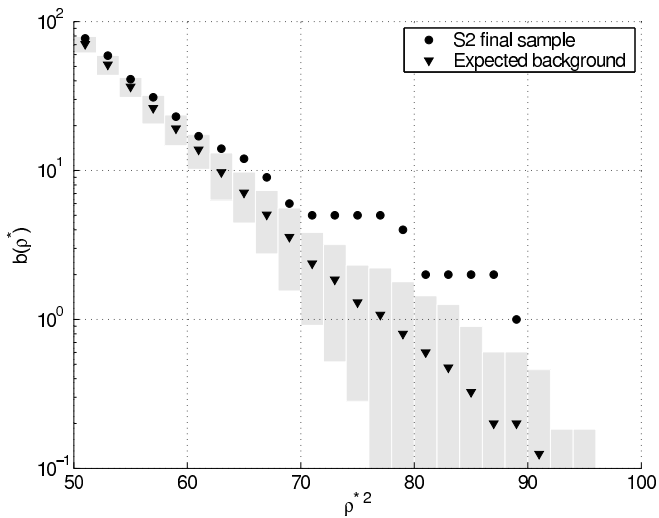}
\end{center}
\caption{\label{f:event-per-s2}
The number of triggers per S2 above combined SNR $\rho^*$.
The triangles represent the expected (mean) background based on 40 
time-shift analyses.  The shaded envelope indicates the sample variance
in the number of events.   The choice of combined SNR in
Eq.~(\protect\ref{e:combined-snr}) is further justified by the
behavior $b(\rho^*)\propto exp(-\rho^{*2})$ down to about $0.3$ events per
S2.  The circles represent the un-shifted inspiral event candidates
(Sec.~\ref{ss:candidates}).}
\end{figure}

Using data from the time-shifts,  we also computed the (sample) mean
number of events per S2 with $\rho > \rho^*$.  This result is shown in
Fig.~\ref{f:event-per-s2};  the shaded bars represent the sample
variance for ease of comparison with the zero-shift distribution.  The
apparent exponential dependence of $b(\rho^*)$ on $\rho^{*2}$ further
supports the choice of combined statistic in Eq.~(\ref{e:combined-snr}).

\section{Search Results}
\label{s:results}

The pipeline described above was used to analyze the S2 data.  The output
of the pipeline is a list of candidate coincident triggers. To decide
whether there is any plausible detection candidate worth following up,
we compare the combined SNR of the candidates with the expected SNR
from the accidental background. If the probability of any candidate
being accidental is small enough, we look at the robustness of the
parameters of the candidate under changes in the pipeline, and we investigate
possible instrumental reasons for these candidates that may have been
overlooked in the initial analysis. 

Independently of whether detection candidates are found, we can use
the results to set an upper limit on the rate of binary neutron star
coalescences per Milky Way Equivalent Galaxy, per year. We use the
same statistics as in the previous search in S1 data~\cite{LIGOS1iul},
measuring the efficiency of the search at the SNR of the loudest
trigger found. We take here the most conservative approach, taking
into account {\em all} triggers at the output of the pipeline: both
those triggers considered to be potential detection candidates {\it
and} those that are consistent with being due to instrumental noise or
consistent with background. We do not include the playground in the
observation time used for calculating the upper limit, since the
playground was used to the tune the pipeline. This is consistent with
our approach of focusing on detection, and not optimizing the pipeline
for upper limit results (which was the reason for not considering
single detector data, for example).

As described in Section~\ref{s:datasample}, after the data quality
cuts, discarding science segments with durations shorter than 2048
sec, and application of the instrumental veto in L1, a total of
373 hours of data were searched for signals, broken up in
double and triple coincidence time as shown in
Figure~\ref{f:S2times}. For the upper limit analysis, including the
background estimation, we only considered the non-playground times,
amounting to 339 hours, of which 65\% (221 hours)
had all three detectors in operation; 26\% (89 hrs) had only
L1 and H1 in operation; and 8.5\% (29 hrs) had only L1 and
H2 in operation.

\subsection{Triggers and event candidates}
\label{ss:candidates}

The output of the pipeline is a list of candidates which are assigned
an SNR according to Eq.~(\ref{e:combined-snr}).  There are
142 candidates in the non-playground final sample with
combined $\rho^2$ greater than 45; the breakdown is 90 candidates
from L1-H2 two detector data, 35 from triple detector data,
and 17 from L1-H1 two detector data. All the candidates in
the triple detector data had SNR in H1 too small to cross the
threshold in H2, and following our pipeline, were accepted as
coincident triggers. Thus, all our coincident triggers are only
double-coincidence candidates.

Table~\ref{t:loudest} lists the ten largest SNR coincident triggers
recorded in the analysis (including the playground). Detailed
investigations of the conspicuous triggers were performed and are
reported below. Nine out of these ten candidates were in the double
detector sample. There was only one (\#10 in the list) which was in
the triple detector data, although it was too weak to require a
corresponding trigger in H2.  Our loudest candidate, as well as five
of our loudest ten, are L1-H2 coincident triggers; in fact, 63\% of
all our candidates are L1-H2 coincident triggers. Since the L1-H2 data
supplies only 9\% of the total observation time, we conclude that the
noise in H2 was significantly different than the noise in H1,
producing more and louder triggers.  For all triggers in
Table~\ref{t:loudest}, the effective distance is larger in the
Livingston detector than in the Hanford detector: although this is
plausible for real signals, as shown in
Fig.~\ref{f:eff_dist_ratio_hist}, it suggests that these candidates
are more likely to originate from instrumental noise.

\begin{table*}
\caption{\label{t:loudest}
The 10 triggers with the largest SNR which remain at the end of the
pipeline.  This table indicates their UTC date,  the time delay between Hanford and Livingston ($\delta t=t_H-t_L$),  the
combined SNR $\rho^2$ (from Eq.~\ref{e:combined-snr}),  the SNR registered in each detector, the value of $\chi^2$ per degree of freedom at each
interferometer (for $2p-2=28$ d.o.f.), the effective distance to an astrophysical event with
the same parameters in each detector, and the binary component masses of the best
matching template (identical for the triggers in both detectors).
The notation L1-Hx means that all three interferometers were in
science mode at the time of this coincident trigger.
}\vspace*{0.1in}
\newdimen\digitwidth\setbox0=\hbox{$0$}\digitwidth=\wd0\catcode`?=\active
\def?{\kern\digitwidth}
\begin{ruledtabular}
\begin{tabular}{ccccccccccccc}
Rank & YYMMDD  & $\delta t$ & Instruments & $\rho^2$ & $\rho_L$ & $\rho_H$ & $\chi^2/\text{d.o.f.}$ & $\chi^2/\text{d.o.f.}$ & $D^L_{\mathrm{eff}}$ & $D^H_{\mathrm{eff}}$  & $m_1$  & $m_2$ \\
& (UTC) & (ms) & & & & & (LLO) & (LHO) &(Mpc) & (Mpc) & ($M_\odot$) & ($M_\odot$) \\
\hline
\# 1 	& 030328 & 	+9.8 	&	 L1-H2&	 89.1 &	 8.9 &	 6.1&	 2.6 &	 3.4&	 2.1 &	 0.9&	 2.23 &	 2.23	 \\ 
\# 2 	& 030327 & 	-1.7 	&	 L1-H2&	 87.1 &	 7.6 &	 11.0&	 2.7 &	 5.1&	 2.1 &	 0.7&	 2.09 &	 2.09	 \\ 
\# 3 	& 030224 & 	-7.6 	&	 L1-H2&	 79.5 &	 6.5 &	 12.3&	 1.2 &	 6.4&	 2.6 &	 0.4&	 3.04 &	 1.19	 \\ 
\# 4 	& 030301 & 	+7.1 	&	 L1-H1&	 79.0 &	 7.7 &	 8.8&	 2.3 &	 4.7&	 2.1 &	 0.6&	 1.45 &	 1.45	 \\ 
\# 5 	& 030224 & 	+5.4 	&	 L1-H1&	 77.2 &	 6.7 &	 11.3&	 1.9 &	 6.5&	 2.0 &	 0.5&	 1.90 &	 1.35	 \\ 
\# 6 	& 030301 & 	+11.0 	&	 L1-H1&	 69.2 &	 6.7 &	 9.9&	 2.0 &	 5.9&	 2.2 &	 0.5&	 1.28 &	 1.28	 \\ 
\# 7 	& 030224 & 	-1.5 	&	 L1-H2&	 67.2 &	 6.9 &	 8.9&	 1.3 &	 4.1&	 2.7 &	 0.8&	 2.46 &	 1.82	 \\ 
\# 8 	& 030328 & 	+4.4 	&	 L1-H2&	 67.1 &	 7.4 &	 7.2&	 2.8 &	 2.3&	 1.9 &	 0.6&	 2.27 &	 1.11	 \\ 
\# 9 	& 030224 & 	-10.5 	&	 L1-H1&	 66.5 &	 6.9 &	 8.8&	 1.6 &	 4.3&	 2.3 &	 0.8&	 3.60 &	 1.20	 \\ 
\# 10 	& 030301 & 	-1.7 	&	 L1-Hx&	 65.9 &	 7.5 &	 6.1&	 2.5 &	 2.7&	 2.5 &	 1.6&	 2.65 &	 2.65	 \\
\end{tabular}
\end{ruledtabular}
\end{table*}

A trigger gets elevated to the status of an event candidate if the
chance occurrence due to noise is small as determined by time-shift
background estimation, calculated as described in
Sec.~\ref{s:background}.  Each event candidate is subjected to follow-up
investigations beyond the level of automation used in our pipeline to
ensure that it is not due to an instrumental or environmental
disturbance.

In Figure~\ref{f:event-per-s2},
a cumulative histogram of the final coincident triggers 
versus $\rho^2$ is overlayed on
the expected background due to accidental coincidences, 
as determined by time-shifts. Even after taking into account
that these are cumulative histograms, so that adjacent bins are
strongly correlated, it appears that the number of coincident
triggers is inconsistent with the expected background for large
$\rho^2$. 
While the origin of this discrepancy is not understood, careful 
examination of
the three coincident triggers with the highest $\rho^2$, detailed below, indicates
that these are not gravitational wave detections.  Moreover, no evidence of
correlated noise between the Livingston and Hanford observatories was
found for times around these triggers. In
Sec.~\ref{ss:modified-clustering}, we demonstrate that a reasonable
modification of our algorithm for clustering multiple coincident
triggers results in good agreement between the coincident trigger
sample and the background estimate.  This shows that the background 
estimate is
not robust with respect to reasonable variations in the analysis
procedure.  

In Sec.~\ref{s:upperlimit}
we derive an upper limit on the rate of inspiral events which is
conservative with respect to any uncertainties in the background
estimate.  Thus, although the discrepancy between the coincident
triggers and the background estimate presented in
Fig.~\ref{f:event-per-s2} is not understood, it does not affect
our ability to detect inspiral signals or set an upper limit on
the inspiral rate.

\subsubsection{Trigger 030328}

The loudest candidate is in a cluster of three coincident triggers in
L1 and H2 on March 28, 2003. The loudest trigger in the cluster, \#~1
in Table~\ref{t:loudest}, has a large SNR in L1 ($\rho=8.9$), but an SNR in
H2 ($\rho=6.1$) which is close to threshold for trigger generation.  We
expect a candidate arising from a real signal to be robust under
small changes in our analysis pipeline. In order to test the
robustness of this particular candidate upon changes in the boundaries
for the chunks used in the analysis, we re-analyzed the data in this
science segment shifting the start time of the chunk by
different amounts in L1 and H2. Although in all cases our analysis
produced similar numbers of triggers in each detector, only in the
original case were there triggers within 11~ms that had identical
masses and were considered candidates.

Other measured parameters of this trigger re-inforce the conclusion
that it should not be promoted to a detection candidate.  The $\chi^2$
per degree of freedom is bad, especially in H2 (3.4); the
$\chi^2$ in L1 is close to our threshold [$\chi^2/(p+\delta^2
\rho^2)=4.6$ compared to the threshold $\chisqthresh=5$].
The effective distances measured by
the detectors are very different, 2.1 Mpc in L1 and 0.9 Mpc in H2.
This ratio of effective distances and the large time delay (10 ms) is
unlikely, though not impossible, for the posited population.
The other triggers in the cluster
(with smaller SNR by definition) have a larger distance ratio, smaller
time delay, smaller $\chi^2$, and different masses in the binary
system (but the same masses in both triggers): $2.9 M_\odot$ and $1.1
M_\odot$.

The time series in the H2 detector does not show anything unusual in
the gravitational wave channel or in other auxiliary channels: this is
consistent with the small SNR measured in H2 for this trigger. The
time series in L1, however, shows several disturbances in a few second
window around the candidate.  Many inspiral triggers are generated at
the time although only three are coincident in mass and time with H2
triggers.  Just 100 ms before the coalescence time of the trigger in
question (and within the duration of the template corresponding to the
trigger), we observed a fast fluctuation in the calibration, less than
62~ms in duration, indicating low optical gain dropping to levels that
would make the feedback loop unstable.  In the 51 minutes when the L1
detector was operational around the time of the candidate, such a low
loop gain happened at three different instances.  Each occurrence
lasted less than 62~ms and was accompanied by a cluster of inspiral
triggers generated by the search code.  Since we averaged the
calibration on a 60-second time scale, these rare fluctuations had not
been observed when considering data quality.  Nevertheless, such a low
gain, even though brief, makes the data and its calibration very
unreliable.  This provides an instrumental reason to veto this trigger
as a detection candidate: the cluster of triggers in L1 containing the
coincident trigger was probably produced by non-linear effects
associated with either the low gain itself or the alignment
fluctuation that produced the low gain in the first place. An
indication of this is that most of the excess power at the time of the
trigger cluster is at sidebands of a narrow line in the spectrum,
corresponding to up-conversion of low-frequency noise to the 120 Hz
electric power line harmonic by some unknown bilinear process.

\subsubsection{Trigger 030327}

The second loudest candidate in Table~\ref{t:loudest} is on March 27,
2003; it is also an L1-H2 trigger, with a combined SNR $\rho^2=87$.
Similar to the loudest candidate, according to the background
estimates, it has a \checkme{20\%} false alarm probability.  Unlike
the loudest candidate, however, this trigger has a smaller SNR in
L1 (7.6) than in H2 (11.0). 

\begin{figure}
\begin{center}
\includegraphics[height=\linewidth,angle=90]{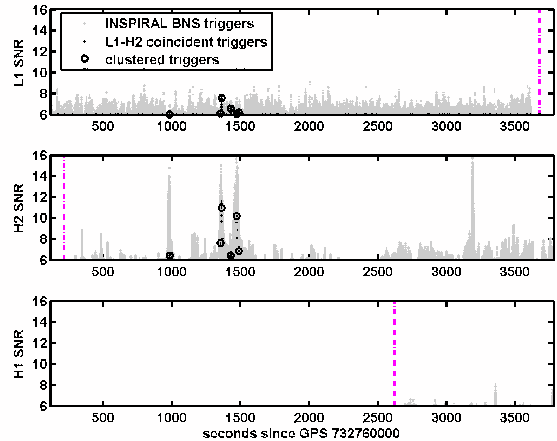}
\caption{Triggers produced by the inspiral search in L1 (top panel),
and the triggered search in H2 (middle panel) and H1 (lower panel),
near the time of our second loudest candidate, on 03/03/27 UTC, in
Table~\ref{t:loudest}. The black points indicate the triggers that
have identical mass and are coincident within a 11 ms window. The
circles indicate the clustered candidate triggers. The dashed lines
are the boundaries of the science segments in each detector (i.e, H1
and H2 were not operating to the left of the dashed line in their
respective graphs, and L1 was not operating after the dashed line in
the top panel) . \label{Trigger030327}}
\end{center}
\end{figure}

As shown in Fig.~\ref{Trigger030327}, 
there are four bursts of noise in the gravitational wave
channel of the H2 detector in the 58 minutes
surrounding this candidate when L1 and H2 were operating in
coincidence.  Each burst is a few to ten seconds long, 
and each triggers a large fraction of the template bank. 
Trigger 030327 is the loudest trigger in a cluster of seven L1-H2
coincident triggers, in a 3~ms interval, during one of the H2 noise
bursts. There are also three other coincident triggers in the same
noise burst, and there are clusters of coincident triggers
in two of the other three H2 noise bursts. One of these bursts,
with several coincident triggers, contains another one of our top twenty
loudest candidates. The only noise burst that does not have any
coincident triggers happens at a time when the H1 detector had begun
operating, and the triple-coincidence criteria were not
satisfied. 
In almost all of the
time-shifts used to estimate the background, there were a number of
L1-H2 coincident triggers at the time of the noise bursts in H2. We
have not found the precise instrumental origin of the disturbances
causing such large numbers of inspiral triggers: although simultaneous
disturbances are observed in a few other auxiliary channels, none of
them are as obvious as they are in the gravitational wave
channel.

Despite a relatively low measured probability of this trigger being
due to the background, we do not consider this trigger an
{event candidate} due to the low SNR in the L1 detector, the
poor $\chi^2$, the unlikely parameters, and a strong suspicion of
instrumental misbehavior.

\subsubsection{Trigger 030224}

We also followed up the third loudest candidate on Feb 24, 2003. This
is again an L1-H2 candidate, happening a few minutes before H2 turns
off due to loss of the operating point at optical resonance. We
discovered a very large feedback loop oscillation in an auxiliary servo
operating in the orthogonal channel to the gravitational channel
(L1:LSC-AS\_I). Although the oscillation was at lower frequencies than
the ones relevant for our search, it produced broadband excess power
at all frequencies in two short time intervals, at the beginning and
at the end of the oscillation. This instrumental misbehavior rules out
this trigger as an event candidate.

\subsection{Background estimation revisited}
\label{ss:modified-clustering}

\begin{figure}
\begin{center}
\includegraphics[width=\linewidth]{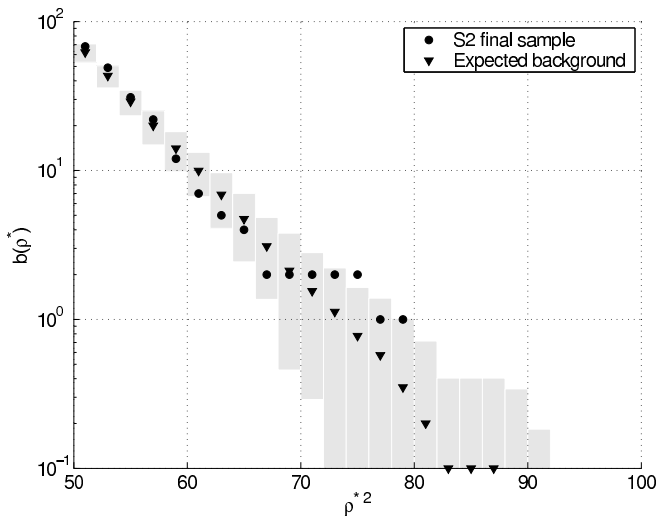}
\end{center}
\caption{\label{f:event-per-s2-new-cluster}
The number of triggers per S2 above combined SNR $\rho^*$ using the
best-fit clustering method.  The triangles represent the expected (mean)
background, while the circles represent zero-shift coincident triggers.
See Fig.~\ref{f:event-per-s2} and Sec.~\ref{s:background}
for details of the time-shifts
and for comparison with largest SNR clustering.  We note that there is
no apparent excess of S2 coincident triggers over the expected background
from accidental coincidences in this plot.}
\end{figure}

One striking feature of our analysis is the apparent discrepancy
between the coincident trigger count in the zero-shift data set compared
to our background estimate shown in Fig.~\ref{f:event-per-s2}.  Our follow-up
investigations rule out gravitational-wave signals as the origin of
this discrepancy. Moreover, no evidence of
correlated noise between the Livingston and Hanford observatories was
found for times around these triggers.  In an effort to understand how
selection effects in our pipeline might be responsible for this
apparent discrepancy,  we repeated our analysis with a different method
of clustering the coincident triggers at the end of our pipeline.
Instead of selecting the trigger with the largest SNR from each
cluster,  we selected the coincident trigger with the minimum value of
\begin{equation}
\frac{\chi^2_{\mathrm{H}}}{p + \delta^2 \rho_{\mathrm{H}}^2} + 
\frac{\chi^2_{\mathrm{L}}}{p + \delta^2 \rho_{\mathrm{L}}^2} \; .
\end{equation}
(Compare to Eq.~\ref{chisqthresh}.)  We call this \emph{best-fit clustering}
since it preferentially selects events with small $\chi^2$ and/or
large SNR. (See Sec.~\ref{ss:analysispipeline} for a description of
the maximum SNR clustering.) The resulting background estimate and
zero-shift distribution are shown in
Fig.~\ref{f:event-per-s2-new-cluster}; no evidence of the original
discrepancy remains.

\begin{figure}
\begin{center}
\includegraphics[width=\linewidth]{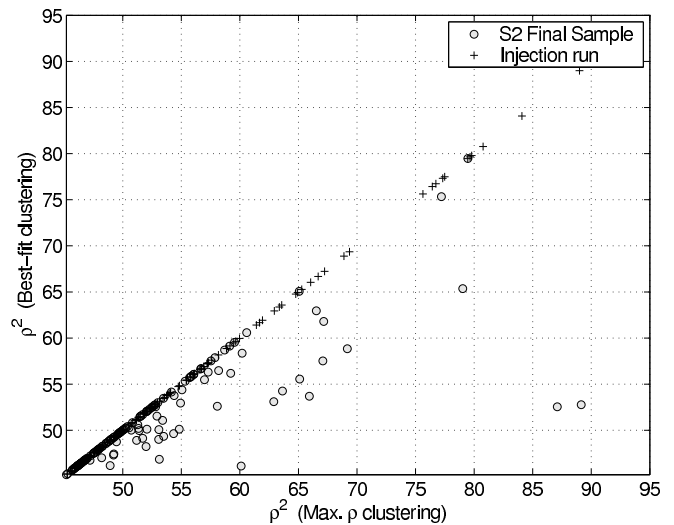}
\end{center}
\caption{\label{f:cluster-compare}A comparison between the SNR
associated with a cluster of triggers by maximizing the SNR over the
cluster and the SNR associated with the same cluster using best-fit
clustering.  Notice that the best-fit clustering often gives lower SNR
values for the final S2 sample, whereas simulated injections have
very similar SNR values for both clustering methods. The triggers 030327,
030328, and 030224 discussed in the text are the grey circles with
horizontal coordinates 89.1, 87.1 and 79.5, respectively.}
\end{figure}

Figure~\ref{f:cluster-compare} compares the SNR under each clustering
scheme for the final S2 trigger sample and a simulated injection run.
The plot shows that the SNR from best-fit clustering is either the
same or
smaller than the corresponding SNR using maximum SNR clustering.
Moreover the values of $\rho^2$ associated with triggers 030328 and
030327, the two loudest triggers in Table~\ref{t:loudest}, 
are less than 55 for best-fit clustering.  
Simulated injections, however, produce
similar values of $\rho^2$ for both clustering methods.  This
observation suggests that best-fit clustering is a better way to
associate an SNR with the triggers than the maximum SNR method.  More
importantly, in our opinion,  it indicates that real signals would be
robust under changes of clustering, thus adding further weight to our
conclusion that the two
loudest triggers in Table \ref{t:loudest}, are not
gravitational waves from inspiral signals. 

\section{Upper limit on the rate of coalescences per galaxy}
\label{s:upperlimit}

Following the notation used in \cite{LIGOS1iul}, let $\mathcal{R}$
indicate the rate of binary neutron star coalescences per year per
Milky Way Equivalent Galaxy (MWEG) and $N_G(\rho^\ast)$ indicate the
number of MWEGs to which our search is sensitive at $\rho \geq
\rho^\ast$.  The probability of observing an inspiral signal with
$\rho > \rho^\ast$ in an observation time $T$ is
\begin{equation}
  P(\rho>\rho^\ast;{\mathcal{R}}) = 1 - e^{-{\mathcal{R}}T N_G(\rho^\ast)}.
  \label{e:foreground-poisson}
\end{equation}
A trigger can arise from either an inspiral signal in the data or from
background.   If $P_b$ denotes the probability that all background triggers
have SNR less than $\rho^\ast$,  then the probability of observing
one or more triggers with $\rho > \rho^\ast$ is given by
\begin{equation}
  P(\rho>\rho^\ast;{\mathcal{R}},b) = 1 - P_b e^{-{\mathcal{R}}TN_G(\rho^\ast)}.
  \label{e:joint-dist}
\end{equation}
Given the probability $P_b$, the total observation time $T$, the 
largest observed signal-to-noise $\rho_{\mathrm{max}}$, and 
the number of MWEGs
$N_{{G}}(\rho_{\mathrm{max}})$ 
to which the search is sensitive, we find that the rate of binary 
neutron star inspirals per MWEG satisfies
\begin{equation}
  {\cal R} < {\cal R}_{90\%} = 
  \frac{2.303+\ln P_b} {T N_{G}(\rho_{\mathrm{max}})}
\end{equation}
with 90\% confidence.  This is a frequentist upper limit on the rate.
For ${\mathcal{R}}>{\mathcal{R}}_{90\%}$, there is more than $90\%$
probability that at least one event would be observed with SNR greater
than $\rho_{\mathrm{max}}$.   Details of this method of determining an 
upper limit can be found in Ref.~\cite{loudestGWDAW03}.  In
particular, one obtains a conservative upper limit by setting 
$P_b=1$;  we adopt this approach below because of uncertainties in our
background estimate.

\begin{figure}
\begin{center}
\includegraphics[width=\linewidth]{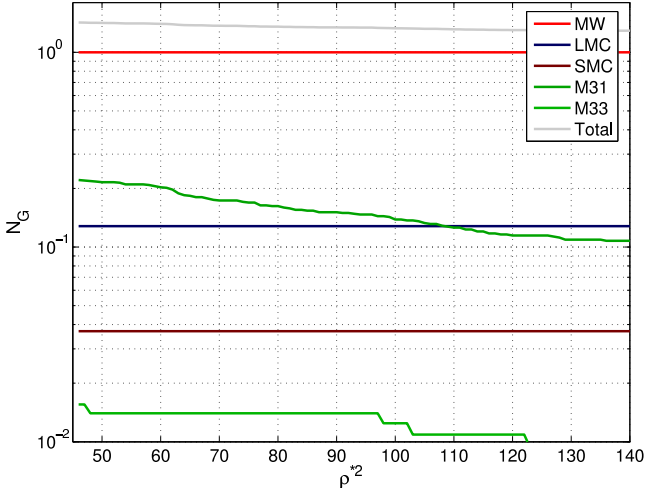}
\end{center}
\caption{\label{f:eff_vs_snrsq}
The efficiency of the search to the target population described in
Sec.~\ref{s:population} as a function of the SNR threshold $\rho^*$.
The contribution of each galaxy to $N_G$ is shown for galaxies which
contribute more than $1\%$ of an MWEG.   The largest SNR of a coincident
trigger observed in this analysis was $\rho^2=89$ meaning that $N_G=\nmweg
\textrm{ MWEG}$ were probed by the search.
}
\end{figure}

During the $T=\checkme{\totaltimehours}\text{ h}$ of data used in our analysis, 
the largest observed SNR was $\rho_{\mathrm{max}}^2
= \checkme{\rhomaxsqr}$.  The number of MWEGs $N_G$ was computed using a Monte
Carlo simulation in which the data was re-analyzed with simulated
inspiral signals, drawn from the population described in
Sec.~\ref{s:population},  added to the time series. The results are
shown in Fig.~\ref{f:eff_vs_snrsq} which breaks down the contribution
to $N_G$ galaxy by galaxy in the target population.  (Results are
shown only for galaxies contributing more than 0.01 MWEG.)
At $\rho^{\ast 2} =\checkme{\rhomaxsqr} $,  we find $N_G = \nmweg \textrm{ MWEG}$;
this is subject to some
uncertainties, to be discussed in the next section.  
We determine that the probability that all background events have
SNR smaller than the largest observed SNR is $P_b=\checkme{0.8}$.  
Since no systematic error has been assigned to this background
estimate, we take $P_b=1$ to be conservative.
As a function of
the true value of $N_{\text{G}}$, the rate limit is
\begin{equation}
\label{eq:limitng}
{\mathcal{R}}_{90\%}=\rpermweg \left( \frac{\nmweg}{N_{\text{G}}}
\right) \text{ y}^{-1}
\text{ MWEG}^{-1} \; .
\end{equation}

\subsection{Error analysis}
\label{s:errors}

The principal systematic effects on our rate limit are (i) inaccuracies in
our model population, including inaccuracies in the inspiral waveform assumed,
and (ii) errors in the calibration of the instrument.  
All other systematic effects in the analysis pipeline (for
example, less-than-perfect coverage of the template bank) are taken
into account by the Monte Carlo estimation of the detection
efficiency.  It is convenient to express the effective number of MWEGs
as
\begin{eqnarray}
N_{\mathrm{G}} &=& \varepsilon(\rho_{\mathrm{max}}) 
\left(\frac{L_{\mathrm{pop}}} {L_{\mathrm{MW}}} \right)
\\
&=&  \sum_{i=0}^n 
\varepsilon_i(\rho_{\mathrm{max}}) \frac{L_i}{L_{MW}}
\end{eqnarray}
where $L_{\mathrm{MW}}$ is the effective blue-light luminosity of the
Milky Way and the index $i$ identifies a galaxy in the population
where $i=0$ corresponds to the Milky Way.  The fraction of the signals
that would be detectable from a particular galaxy in this search is
denoted $\varepsilon_i$ -- we will refer to this as the efficiency of
the search to sources in galaxy $i$. Referring to
Table~\ref{t:galaxies}, $\varepsilon_i(\rho_{\mathrm{max}})$ is given
in column 6, $L_i/L_{\mathrm{MW}}$ in column 5, and the cumulative
value of $N_{\mathrm{G}}$ in column 7.

\subsubsection{Uncertainties in population model}
\label{ss:popuncertainties}
Uncertainties in the population model used for the Monte-Carlo
simulations may lead to differences between the inferred rate and the
rate in the universe. Since the effective blue-light luminosity
$L_{\text{pop}}$ is normalized to our Galaxy,   variations arise
from the relative contributions of other galaxies in the population.
These contributions depend on the estimated distances to the galaxies,  
estimated reddening,  and corrections
for metallicity (lower values tend to produce higher mass binaries),
among other things.  

The spatial distribution of the sources can also introduce significant
uncertainties. Typically, the distances to nearby galaxies are only known
to about 10\% accuracy.  In fact, the distance to Andromeda is thought
to suffer from a $15\%$ error.  A change in distance $\Delta d$ to a galaxy
assumed to be at distance $d$ introduces two different uncertainties in 
$N_{\mathrm{G}}$:
\begin{enumerate}
\item The change in our estimate of the efficiency 
for the particular galaxy is determined
by observing that the correct signal to noise ratio decreases
inversely with the distance to the galaxy and hence
\begin{equation}
\varepsilon_i(\rho_{\mathrm{max}})
\rightarrow
\varepsilon_i(\rho_{\mathrm{max}} [1 + \Delta d / d])
\end{equation}
so that efficiency decreases as the distance increases.  

\item The change in our estimate of the luminosity that is correlated with this change in
distance is given by 
\begin{equation}
\frac{L_i}{L_{\mathrm{MW}}}
\rightarrow
\frac{L_i}{L_{\mathrm{MW}}} 
\left( 
\frac{d + \Delta d}{d}
\right)^2
\end{equation}
so that our estimate of the luminosity increases as the distance increases.
\end{enumerate}
Adopting
a $15\%$ error in all distances,  we estimate the error in $N_G$ due
to uncertainties in galactic distances to be
\begin{equation}
\left. \delta N_G \right|_{d} = 0.04 \textrm{ MWEG} \; .
\end{equation}

The absolute blue light luminosity of the Milky Way is uncertain since
it is inferred from other galactic parameters.  There appears to be
some confusion about the this in the literature on binary neutron star
rate estimates.  Phinney~\cite{Phinney:1991ei} used $L_{\mathrm{MW}} =
1.6 \times 10^{10} L_{\odot, B}$ while the more recent work of
Kalogera, Narayan, Spergel and Taylor~\cite{Kalogera:2000dz} uses $9
\times 10^9 L_{\odot,V}$ citing a factor of two difference with the
number used by Phinney;  these number agree within 1\%, however, once 
converted to the same units.  

Errors in the blue magnitude of each galaxy and relative normalization
to the Milky Way enter in the same way.   For the galaxies that
contribute most substantially to $N_G$,   the LEDA catalog suggests
errors of $\sim 10\%$ in corrected blue magnitude which translates into errors of
$\sim 20\%$ in luminosity.   Adopting this uniformly for the galaxies
in our population,   we estimate
\begin{equation}
 \left. \delta N_G \right|_L = 0.04 \textrm{ MWEG} 
\end{equation}

Different models for NS-NS formation can lead to variations in
the NS mass distribution~\cite{Belczynski:2002}, although the bulk of the
distribution always remains strongly peaked around observed NS
masses~\cite{Thorsett:1998uc}.   We estimate the corresponding
variations in $N_G$ to be
\begin{equation}
\left. \delta N_G
\right|_{\mathrm{mass}} \approx 0.01 \textrm{ MWEG}
\end{equation}
based on simulations with 50\% reduction in the number of binary systems
with masses in the range $1.5 M_\odot < m_1,m_2 < 3.0 M_\odot$.  This
is indicative only. A few extreme scenarios for binary formation can produce
more severe alterations to the mass distribution~\cite{Belczynski:2002}.

The waveforms used both in the Monte-Carlo simulation and in the
detection templates ignore spin effects.  Estimates based on the work
of Apostolatos~\cite{Apostolatos:1995} suggest that less than 10\% of
all spin-orientations and parameters consistent with binary neutron
stars provide a loss of SNR greater than $\sim 5\%$.  Thus, the
mismatch between the signal from spinning neutron stars and our
templates should not significantly affect the upper limit.  To be
conservative, however, we place an upper limit only on non-spinning
neutron stars; we will address this issue quantatively in future
analysis.

\subsubsection{Uncertainties in the instrumental response}
\label{ss:respuncertainties}

The uncertainty in the calibration can be quantified as an uncertainty
in amplitude and phase of a frequency dependent response function.

Using the same procedure as in S1~\cite{Gonzalez:2002}, the average 
instrument response $R(f)$ was
constructed for every minute of data during S2 from a reference
sensing function $C(f)$, a reference open loop gain function $G(f)$,
and a parameter $\alpha(t)$ representing varying optical gain. The
response function at a time $t$ during the run is given by
$R(f,t)=(1+\alpha(t)G(f))/(\alpha(t)C(f))$. The parameter $\alpha$ was
reconstructed using the observed amplitude of a calibration line.  If
an inspiral signal is present in the data, errors in the
calibration can cause a mismatch between the template and the signal.
We can quantify and measure this effect using simulated injections,
both in software and hardware. For such injections, the SNR differs
from the SNR that would be recorded for a signal from a real inspiral
event at the same distance as the injection.  The effect is linear in
amplitude errors causing either an upward or downward shift in SNR; 
in contrast, the effect is quadratic in phase errors causing an 
over-estimation of sensitivity~\cite{Brown:2004}.  This error is 
propagated to our upper
limit by shifting the efficiency curve in Fig.~\ref{f:eff_vs_snrsq}
horizontally by the appropriate amount.

A careful evaluation of uncertainties in the S2
calibration~\cite{Gonzalez:2004} has shown that amplitude errors have
two dominant components. The first source of error, an imperfect
knowledge of the strength of the feedback actuators in the system,
produces an error in the overall amplitude of the sensing function
$C(f)$. Thus, it manifests as a systematic error in the amplitude of
the response function, constant during the whole run. This error was
estimated to be 8.5\% in L1, 3.5\% in H1, and 4.5\% in H2.

The second dominant source of error is an imprecise measurement of the
amplitude of the calibration line, resulting in an error of the
coefficient $\alpha(t)$. This measurement error is mostly random in
nature, and leads to magnitude and phase errors in the response
function. 
These errors translate into maximum amplitude errors in the response
function equal to 6\% for L1 and H2, and  18\% for H1. 

Another source of error in $\alpha(t)$, which is not captured in the
errors as estimated above, is due to changes in optical gain that are
averaged over the 60 second integration time used in the measurement
of $\alpha(t)$. Based on a limited set of diagnostics, we roughly
estimate these errors can be as large as 10\% in L1, and smaller in H1
and H2. The errors in $\alpha$, including these fast fluctuations, are
random in nature, and are not expected to contribute to the error in
the measured efficiency.

The SNR used in the final analysis is constructed from the individual
SNR's $\rho_L$ and $\rho_H$ as given in Eq.~(\ref{e:combined-snr}).
The error in SNR then has two pieces 
\begin{eqnarray}
(\delta \rho )^2 &=& \frac{\rho_L^2}{\rho^2} (\delta \rho_L )^2 +
        \frac{1}{4^2} \frac{\rho_H^2}{\rho^2} (\delta \rho_H )^2 \\
        &\leq& (\delta \rho_L )^2 + \frac{1}{4^2} (\delta \rho_H )^2
\end{eqnarray}
We assume fractional errors equal to the maximum errors in
calibration at each site and arising from the systematic uncertainty in
the detector response (8.5\% in L1, and 4.5\% for LHO). This results
in a conservative estimate of the error at the largest observed SNR:
\begin{equation}        
   (\delta \rho )^2      
        \leq (0.085 \rho_{\mathrm{max}})^2 + 0.0625 (0.045 \rho_{\mathrm{max}} )^2
        \; .
\end{equation}
The resulting error on $N_G$ is
\begin{equation}
\left. \delta N_G \right|_{\mathrm{cal}} = 0.02 \textrm{ MWEG} 
\end{equation}

Simulations of the contributions to $\delta N_G$ from the random
fluctuations of $\alpha$, assuming the largest possible error in H1
(18\%), made negligible contributions to the error due to calibration,
as expected.

\subsubsection{Uncertainties in the analysis pipeline}
\label{ss:pipuncertainties}

Since we use matched filtering to search for gravitational waves from
inspiralling binaries, differences between the theoretical and the
real waveforms could also adversely effect the results.  These effects
have been studied in great detail for binary neutron star
systems~\cite{Apostolatos:1995, Droz:1997, Droz:1998}.  The results indicate
a
$\sim 10\%$ loss of SNR due to inaccurate modelling of the waveforms
for binaries in the mass range of interest.  This feeds into our
result through our measurement of the efficiency.  We may be
over-estimating our sensitivity to real binary inspiral signals;
the estimated loss of efficiency results in an error on $N_{G}$ of
\begin{equation}
\left. \delta N_G \right|_{\mathrm{wave}} = +0/-0.02 \textrm{ MWEG} 
\end{equation}

The effects of discreteness of the template placement,  errors in the
estimates of the power spectral density $S(f)$ used in the matched
filter in Eq.~(\ref{e:xfilter}),  and trends in the instrumental noise
are all accounted for by the Monte-Carlo simulation.   The error in
the efficiency measurement due to the finite number of injections is
estimated as 
\begin{equation}
\left. \delta N_G^i \right|_{\mathrm{MC}} = L_i \sqrt{ \varepsilon_i
(1-\varepsilon_i) / n^i}
\end{equation}
where $n^i$ is the number of injections made for each galaxy.   
The total error from the Monte Carlo
is
\begin{equation}
\left. \delta N_G \right|_{\mathrm{MC}} = 0.02 \textrm{ MWEG} 
\end{equation}

\subsubsection{Combined uncertainties on $N_{\text{G}}$ and the rate}
\label{ss:combuncertainties}

Combining the errors in quadrature
yields total errors 
\begin{equation}
\left. \delta N_G \right|_{\mathrm{total}} = \plusminus{0.06}{0.07}
\textrm{ MWEG} 
\end{equation}
To be conservative, we assume the downward excursion $N_{\mathrm{G}} =
\nmweg - 0.07 = 1.27 \textrm{ MWEG}$ when using
Eq.~(\ref{eq:limitng}) to derive an observational upper limit on the
rate of binary neutron star coalescence
\begin{equation}
{\mathcal{R}} < \ratepermweg \textrm{ y}^{-1} \textrm{ MWEG}^{-1}
\end{equation}
where we have used $T = \totaltimehours \textrm{ hr}$ and $P_b=1$.

\section{Summary and Discussion}
\label{s:conclusion}

Using data from the second LIGO science run (S2), we have significantly
improved our methods and strategies to search for waveforms from
inspiraling neutron stars. The search has been optimized for detecting
signals, rather than for setting optimal upper limits on the rate of
sources~\cite{LIGOS1iul}. In order to increase confidence in a detection
candidate, we used data
only when two or more detectors are operating.  We performed
extensive validations of the detection efficiency of our search using
software and hardware signal injections. Using these injections, we
tuned the parameters in our search to take advantage of the
increased reach of the detectors in the S2 run and have almost 100\%
efficiency for signals in the Milky Way and the Magellanic clouds.
(Due to the antenna pattern,  we cannot be 100\% efficient.)
We also
achieved an average 6\% efficiency for sources in the Andromeda galaxy, at a
distance of 770 kpc, near the typical maximum range of our second best
detector (H1, 900 kpc). Using the same detection pipeline, but
shifting in time the data from the two observatories, we
quantified the accidental rate for our pipeline.  This allows 
evaluation of our confidence that any candidate is a true
signal.

The search of 373 hours of double- and triple-detector
data during S2 resulted in no detection: the strongest candidate
from the search, arising from coincidence between L1 and H2, had a
combined signal-to-noise of 9.4.  Our background estimate suggests
that noise has a \checkme{20\%} probability of causing a candidate
with this signal-to-noise or greater.  Moreover,  the weight of
evidence from follow-up investigations suggests that this trigger
originated with a burst of noise which corresponded to a brief period
of servo instability.    The detailed investigations presented in
Sec.~\ref{s:results} exemplify typical follow-up of 
coincident triggers which appear unlikely to arise from background.  We have
found no evidence of a gravitational wave event from
binary neutron star inspiral.  Without a detection, the 339 hours of
non-playground data were used to place an upper limit on the rate of
binary neutron star coalescence in the Universe.

This search was sensitive to binary inspiral signals from neutron star
coalescence in the Local Group and other galaxies at distances of up
to about 1.5~Mpc. We used Monte Carlo simulations to measure
the efficiency of our search to non-spinning neutron stars in
this population.  We
conclude that the rate of binary neutron star coalescence is
$\cal{R}\leq {\ratepermweg}~{\rm yr^{-1} MWEG^{-1}}$ with 90\%
confidence. This rate limit is significantly greater than
astrophysically plausible rates~\cite{Kalogera:2004tn}, but it
illustrates the performance of the search.  It should be noted that
the analysis of the first science run~\cite{LIGOS1iul} was optimized
to yield the lowest upper limit, while the method presented here was
not.  In particular, the decision to only use data when the Livingston
interferometer {\it and} one of the Hanford interferometers were
operating limited the observation time to 1/3 of the total data
collected. Analysis of the remaining data collected in coincidence with TAMA will be reported elsewhere~\cite{ligotama}.

In this paper,  we have presented a data analysis strategy that could
lead to a detection of gravitational waves from binary neutron star
inspirals. The methods used to validate the search illustrate the
subtleties of the analysis of several detectors with different
sensitivities and orientations.  Moreover,  the experience gained by
following up the largest coincident triggers will be crucial input to
investigations of event candidates that are identified in future
searches.

\acknowledgments

The authors gratefully acknowledge the support of the United States
National Science Foundation for the construction and operation of the
LIGO Laboratory and the Particle Physics and Astronomy Research
Council of the United Kingdom, the Max-Planck-Society and the State of
Niedersachsen/Germany for support of the construction and operation of
the GEO600 detector. The authors also gratefully acknowledge the
support of the research by these agencies and by the Australian
Research Council, the Natural Sciences and Engineering Research
Council of Canada, the Council of Scientific and Industrial Research
of India, the Department of Science and Technology of India, the
Spanish Ministerio de Educaci{\'o}n y Ciencia, the John Simon Guggenheim
Foundation, the Leverhulme Trust, the David and Lucile Packard
Foundation, the Research Corporation, and the Alfred P. Sloan
Foundation.


\begin{thebibliography}{40}
\expandafter\ifx\csname natexlab\endcsname\relax\def\natexlab#1{#1}\fi
\expandafter\ifx\csname bibnamefont\endcsname\relax
  \def\bibnamefont#1{#1}\fi
\expandafter\ifx\csname bibfnamefont\endcsname\relax
  \def\bibfnamefont#1{#1}\fi
\expandafter\ifx\csname citenamefont\endcsname\relax
  \def\citenamefont#1{#1}\fi
\expandafter\ifx\csname url\endcsname\relax
  \def\url#1{\texttt{#1}}\fi
\expandafter\ifx\csname urlprefix\endcsname\relax\def\urlprefix{URL }\fi
\providecommand{\bibinfo}[2]{#2}
\providecommand{\eprint}[2][]{\url{#2}}

\bibitem[{\citenamefont{Saulson}(1994)}]{Saulson:1994}
\bibinfo{author}{\bibfnamefont{P.~R.} \bibnamefont{Saulson}},
  \emph{\bibinfo{title}{Fundamentals of Interferometric Gravitational Wave
  Detectors}} (\bibinfo{publisher}{World Scientific},
  \bibinfo{address}{Singapore}, \bibinfo{year}{1994}).

\bibitem[{\citenamefont{Barish and Weiss}(1999)}]{Barish:1999}
\bibinfo{author}{\bibfnamefont{B.~C.} \bibnamefont{Barish}} \bibnamefont{and}
  \bibinfo{author}{\bibfnamefont{R.}~\bibnamefont{Weiss}},
  \bibinfo{journal}{Phys.\ Today} \textbf{\bibinfo{volume}{52 (Oct)}},
  \bibinfo{pages}{44} (\bibinfo{year}{1999}).

\bibitem[{\citenamefont{Abbott et~al.}(2004{\natexlab{a}})}]{LIGOS1instpaper}
\bibinfo{author}{\bibfnamefont{B.}~\bibnamefont{Abbott}} \bibnamefont{et~al.}
  (\bibinfo{collaboration}{LIGO Scientific Collaboration}),
  \bibinfo{journal}{Nucl. Instrum. Methods} \textbf{\bibinfo{volume}{A517}},
  \bibinfo{pages}{154} (\bibinfo{year}{2004}{\natexlab{a}}).

\bibitem[{\citenamefont{Cutler et~al.}(1993)}]{Cutler:1992tc}
\bibinfo{author}{\bibfnamefont{C.}~\bibnamefont{Cutler}} \bibnamefont{et~al.},
  \bibinfo{journal}{Phys. Rev. Lett.} \textbf{\bibinfo{volume}{70}},
  \bibinfo{pages}{2984} (\bibinfo{year}{1993}).

\bibitem[{\citenamefont{Abbott et~al.}(2004{\natexlab{b}})}]{LIGOS1iul}
\bibinfo{author}{\bibfnamefont{B.}~\bibnamefont{Abbott}} \bibnamefont{et~al.}
  (\bibinfo{collaboration}{LIGO Scientific Collaboration}),
  \bibinfo{journal}{Phys. Rev. D} \textbf{\bibinfo{volume}{69}},
  \bibinfo{pages}{122001} (\bibinfo{year}{2004}{\natexlab{b}}).

\bibitem[{\citenamefont{Kalogera et~al.}(2004)}]{Kalogera:2004tn}
\bibinfo{author}{\bibfnamefont{V.}~\bibnamefont{Kalogera}}
  \bibnamefont{et~al.}, \bibinfo{journal}{Astrophys. J.}
  \textbf{\bibinfo{volume}{601}}, \bibinfo{pages}{L179} (\bibinfo{year}{2004}).

\bibitem[{\citenamefont{Abbott et~al.}(2005{\natexlab{a}})}]{LIGOS2grb}
\bibinfo{author}{\bibfnamefont{B.}~\bibnamefont{Abbott}} \bibnamefont{et~al.}
  (\bibinfo{collaboration}{LIGO Scientific Collaboration})
  (\bibinfo{year}{2005}{\natexlab{a}}), \eprint{gr-qc/0501068}.

\bibitem[{\citenamefont{Ando~et al.}(2001)}]{TAMA:2001}
\bibinfo{author}{\bibfnamefont{M.}~\bibnamefont{Ando~et al.}},
  \bibinfo{journal}{Phys. Rev. Lett.} \textbf{\bibinfo{volume}{86}},
  \bibinfo{pages}{3950} (\bibinfo{year}{2001}).

\bibitem[{\citenamefont{Stairs}(2004)}]{Stairs:2004}
\bibinfo{author}{\bibfnamefont{I.~H.} \bibnamefont{Stairs}},
  \bibinfo{journal}{Science} \textbf{\bibinfo{volume}{304}},
  \bibinfo{pages}{547} (\bibinfo{year}{2004}).

\bibitem[{\citenamefont{Weisberg and Taylor}(2003)}]{weisberg:2002}
\bibinfo{author}{\bibfnamefont{J.~M.} \bibnamefont{Weisberg}} \bibnamefont{and}
  \bibinfo{author}{\bibfnamefont{J.~H.} \bibnamefont{Taylor}}, in
  \emph{\bibinfo{booktitle}{Radio Pulsars}}, edited by
  \bibinfo{editor}{\bibfnamefont{M.}~\bibnamefont{Bailes}},
  \bibinfo{editor}{\bibfnamefont{D.~J.} \bibnamefont{Nice}}, \bibnamefont{and}
  \bibinfo{editor}{\bibfnamefont{S.}~\bibnamefont{Thorsett}}
  (\bibinfo{publisher}{ASP. Conf. Series}, \bibinfo{year}{2003}).

\bibitem[{\citenamefont{Phinney}(1991)}]{Phinney:1991ei}
\bibinfo{author}{\bibfnamefont{E.~S.} \bibnamefont{Phinney}},
  \bibinfo{journal}{Astrophys. J.} \textbf{\bibinfo{volume}{380}},
  \bibinfo{pages}{L17} (\bibinfo{year}{1991}).

\bibitem[{\citenamefont{Belczynski et~al.}(2002)\citenamefont{Belczynski,
  Kalogera, and Bulik}}]{Belczynski:2002}
\bibinfo{author}{\bibfnamefont{K.}~\bibnamefont{Belczynski}},
  \bibinfo{author}{\bibfnamefont{V.}~\bibnamefont{Kalogera}}, \bibnamefont{and}
  \bibinfo{author}{\bibfnamefont{T.}~\bibnamefont{Bulik}},
  \bibinfo{journal}{Astrophys. J.} \textbf{\bibinfo{volume}{572}},
  \bibinfo{pages}{407} (\bibinfo{year}{2002}).

\bibitem[{\citenamefont{Abbott et~al.}(2005{\natexlab{b}})}]{LIGOs2macho}
\bibinfo{author}{\bibfnamefont{B.}~\bibnamefont{Abbott}} \bibnamefont{et~al.}
  (\bibinfo{collaboration}{LIGO Scientific Collaboration}),
  \bibinfo{journal}{submitted to Phys.~Rev. D}
  (\bibinfo{year}{2005}{\natexlab{b}}).

\bibitem[{\citenamefont{Mateo}(1998)}]{Mateo:1998}
\bibinfo{author}{\bibfnamefont{M.}~\bibnamefont{Mateo}},
  \bibinfo{journal}{Annu. Rev. Astron. Astrophys.}
  \textbf{\bibinfo{volume}{36}}, \bibinfo{pages}{435} (\bibinfo{year}{1998}).

\bibitem[{\citenamefont{{Tully}}(1994)}]{Tully:1994}
\bibinfo{author}{\bibfnamefont{R.~B.} \bibnamefont{{Tully}}},
  \bibinfo{journal}{VizieR Online Data Catalog}
  \textbf{\bibinfo{volume}{7145}}, \bibinfo{pages}{0} (\bibinfo{year}{1994}).

\bibitem[{\citenamefont{de~Vaucouleurs
  et~al.}(1991)\citenamefont{de~Vaucouleurs, de~Vaucouleurs, Corwin, Buta,
  Paturel, and Fouque}}]{RC3:1991}
\bibinfo{author}{\bibfnamefont{G.}~\bibnamefont{de~Vaucouleurs}},
  \bibinfo{author}{\bibfnamefont{A.}~\bibnamefont{de~Vaucouleurs}},
  \bibinfo{author}{\bibfnamefont{H.~G.} \bibnamefont{Corwin}},
  \bibinfo{author}{\bibfnamefont{R.~J.} \bibnamefont{Buta}},
  \bibinfo{author}{\bibfnamefont{G.}~\bibnamefont{Paturel}}, \bibnamefont{and}
  \bibinfo{author}{\bibfnamefont{P.}~\bibnamefont{Fouque}},
  \emph{\bibinfo{title}{Third Reference Catalogue of Bright Galaxies (RC3)}}
  (\bibinfo{publisher}{Springer-Verlag:}, \bibinfo{address}{New York, NY},
  \bibinfo{year}{1991}).

\bibitem[{\citenamefont{Nutzman et~al.}(2004)\citenamefont{Nutzman, Kalogera,
  Finn, Hendrickson, and Belczynski}}]{Nutzman:2004}
\bibinfo{author}{\bibfnamefont{P.}~\bibnamefont{Nutzman}},
  \bibinfo{author}{\bibfnamefont{V.}~\bibnamefont{Kalogera}},
  \bibinfo{author}{\bibfnamefont{L.~S.} \bibnamefont{Finn}},
  \bibinfo{author}{\bibfnamefont{C.}~\bibnamefont{Hendrickson}},
  \bibnamefont{and}
  \bibinfo{author}{\bibfnamefont{K.}~\bibnamefont{Belczynski}},
  \bibinfo{journal}{Astrophys. J.} \textbf{\bibinfo{volume}{612}},
  \bibinfo{pages}{364} (\bibinfo{year}{2004}).

\bibitem[{\citenamefont{Apostolatos}(1995)}]{Apostolatos:1995}
\bibinfo{author}{\bibfnamefont{T.~A.} \bibnamefont{Apostolatos}},
  \bibinfo{journal}{Phys. Rev. D} \textbf{\bibinfo{volume}{52}},
  \bibinfo{pages}{605} (\bibinfo{year}{1995}).

\bibitem[{\citenamefont{Droz and Poisson}(1997)}]{Droz:1997}
\bibinfo{author}{\bibfnamefont{S.}~\bibnamefont{Droz}} \bibnamefont{and}
  \bibinfo{author}{\bibfnamefont{E.}~\bibnamefont{Poisson}},
  \bibinfo{journal}{Phys. Rev. D} \textbf{\bibinfo{volume}{56}},
  \bibinfo{pages}{4449} (\bibinfo{year}{1997}).

\bibitem[{\citenamefont{Droz}(1999)}]{Droz:1998}
\bibinfo{author}{\bibfnamefont{S.}~\bibnamefont{Droz}}, \bibinfo{journal}{Phys.
  Rev. D} \textbf{\bibinfo{volume}{59}}, \bibinfo{pages}{064030}
  (\bibinfo{year}{1999}).

\bibitem[{\citenamefont{Bildsten and Cutler}(1992)}]{Bildsten:1992my}
\bibinfo{author}{\bibfnamefont{L.}~\bibnamefont{Bildsten}} \bibnamefont{and}
  \bibinfo{author}{\bibfnamefont{C.}~\bibnamefont{Cutler}},
  \bibinfo{journal}{Astrophys. J.} \textbf{\bibinfo{volume}{400}},
  \bibinfo{pages}{175} (\bibinfo{year}{1992}).

\bibitem[{\citenamefont{Blanchet et~al.}(1996)\citenamefont{Blanchet, Iyer,
  Will, and Wiseman}}]{Blanchet:1996pi}
\bibinfo{author}{\bibfnamefont{L.}~\bibnamefont{Blanchet}},
  \bibinfo{author}{\bibfnamefont{B.~R.} \bibnamefont{Iyer}},
  \bibinfo{author}{\bibfnamefont{C.~M.} \bibnamefont{Will}}, \bibnamefont{and}
  \bibinfo{author}{\bibfnamefont{A.~G.} \bibnamefont{Wiseman}},
  \bibinfo{journal}{Class. Quant. Grav.} \textbf{\bibinfo{volume}{13}},
  \bibinfo{pages}{575} (\bibinfo{year}{1996}).

\bibitem[{\citenamefont{Droz et~al.}(1999)\citenamefont{Droz, Knapp, Poisson,
  and Owen}}]{Droz:1999qx}
\bibinfo{author}{\bibfnamefont{S.}~\bibnamefont{Droz}},
  \bibinfo{author}{\bibfnamefont{D.~J.} \bibnamefont{Knapp}},
  \bibinfo{author}{\bibfnamefont{E.}~\bibnamefont{Poisson}}, \bibnamefont{and}
  \bibinfo{author}{\bibfnamefont{B.~J.} \bibnamefont{Owen}},
  \bibinfo{journal}{Phys. Rev. D} \textbf{\bibinfo{volume}{59}},
  \bibinfo{pages}{124016} (\bibinfo{year}{1999}).

\bibitem[{\citenamefont{Owen}(1996)}]{Owen:1995tm}
\bibinfo{author}{\bibfnamefont{B.~J.} \bibnamefont{Owen}},
  \bibinfo{journal}{Phys. Rev. D} \textbf{\bibinfo{volume}{53}},
  \bibinfo{pages}{6749} (\bibinfo{year}{1996}).

\bibitem[{\citenamefont{Owen and Sathyaprakash}(1999)}]{Owen:1998dk}
\bibinfo{author}{\bibfnamefont{B.~J.} \bibnamefont{Owen}} \bibnamefont{and}
  \bibinfo{author}{\bibfnamefont{B.~S.} \bibnamefont{Sathyaprakash}},
  \bibinfo{journal}{Phys. Rev. D} \textbf{\bibinfo{volume}{60}},
  \bibinfo{pages}{022002} (\bibinfo{year}{1999}).

\bibitem[{\citenamefont{Allen}(2005)}]{Allen:2004}
\bibinfo{author}{\bibfnamefont{B.}~\bibnamefont{Allen}},
  \bibinfo{journal}{Phys. Rev. D} \textbf{\bibinfo{volume}{71}},
  \bibinfo{pages}{062001} (\bibinfo{year}{2005}).

\bibitem[{\citenamefont{Ito}()}]{glitchMon}
\bibinfo{author}{\bibfnamefont{M.}~\bibnamefont{Ito}},
  \emph{\bibinfo{title}{glitchmon: A {DMT} monitor to look for transient
  signals in selected channels}}, \bibinfo{note}{developed using the LIGO Data
  Monitoring Tool (DMT) library}.

\bibitem[{\citenamefont{Christensen
  et~al.}(2004{\natexlab{a}})\citenamefont{Christensen, Shawhan, and
  Gonz{\'a}lez}}]{vetoGWDAW03}
\bibinfo{author}{\bibfnamefont{N.}~\bibnamefont{Christensen}},
  \bibinfo{author}{\bibfnamefont{P.}~\bibnamefont{Shawhan}}, \bibnamefont{and}
  \bibinfo{author}{\bibfnamefont{G.}~\bibnamefont{Gonz{\'a}lez}},
  \bibinfo{journal}{Class. Quant. Grav.} \textbf{\bibinfo{volume}{21}},
  \bibinfo{pages}{S1747} (\bibinfo{year}{2004}{\natexlab{a}}).

\bibitem[{LAL()}]{LALS2BNS}
\emph{\bibinfo{title}{{\normalfont LSC Algorithm Library software packages
  {\scshape lal} and {\scshape lalapps}}}}, \bibinfo{note}{the CVS tag versions
  \texttt{iulgroup\_20040512} of \textsc{lal} and \textsc{lalapps} were used in
  this analysis.}, \urlprefix\url{http://www.lsc-group.phys.uwm.edu/lal}.

\bibitem[{\citenamefont{Tannenbaum et~al.}(2001)\citenamefont{Tannenbaum,
  Wright, Miller, and Livny}}]{beowulfbook-condor}
\bibinfo{author}{\bibfnamefont{T.}~\bibnamefont{Tannenbaum}},
  \bibinfo{author}{\bibfnamefont{D.}~\bibnamefont{Wright}},
  \bibinfo{author}{\bibfnamefont{K.}~\bibnamefont{Miller}}, \bibnamefont{and}
  \bibinfo{author}{\bibfnamefont{M.}~\bibnamefont{Livny}}, in
  \emph{\bibinfo{booktitle}{Beowulf Cluster Computing with {L}inux}}, edited by
  \bibinfo{editor}{\bibfnamefont{T.}~\bibnamefont{Sterling}}
  (\bibinfo{publisher}{MIT Press}, \bibinfo{year}{2001}).

\bibitem[{\citenamefont{Fairhurst et~al.}(2004)}]{S2Hardware:2004}
\bibinfo{author}{\bibfnamefont{S.}~\bibnamefont{Fairhurst}}
  \bibnamefont{et~al.}, \emph{\bibinfo{title}{Analysis of inspiral hardware
  injections during s2}} (\bibinfo{year}{2004}), \bibinfo{note}{in
  preparation}.

\bibitem[{\citenamefont{Christensen
  et~al.}(2004{\natexlab{b}})\citenamefont{Christensen, Meyer, and
  Libson}}]{MCMC:2004}
\bibinfo{author}{\bibfnamefont{N.}~\bibnamefont{Christensen}},
  \bibinfo{author}{\bibfnamefont{R.}~\bibnamefont{Meyer}}, \bibnamefont{and}
  \bibinfo{author}{\bibfnamefont{A.}~\bibnamefont{Libson}},
  \bibinfo{journal}{Class. Quant. Grav.} \textbf{\bibinfo{volume}{21}},
  \bibinfo{pages}{317} (\bibinfo{year}{2004}{\natexlab{b}}).

\bibitem[{\citenamefont{Pai et~al.}(2001)\citenamefont{Pai, Dhurandhar, and
  Bose}}]{Pai:2000zt}
\bibinfo{author}{\bibfnamefont{A.}~\bibnamefont{Pai}},
  \bibinfo{author}{\bibfnamefont{S.}~\bibnamefont{Dhurandhar}},
  \bibnamefont{and} \bibinfo{author}{\bibfnamefont{S.}~\bibnamefont{Bose}},
  \bibinfo{journal}{Phys. Rev. D} \textbf{\bibinfo{volume}{64}},
  \bibinfo{pages}{042004} (\bibinfo{year}{2001}).

\bibitem[{\citenamefont{Brady et~al.}(2004)\citenamefont{Brady, Creighton, and
  Wiseman}}]{loudestGWDAW03}
\bibinfo{author}{\bibfnamefont{P.~R.} \bibnamefont{Brady}},
  \bibinfo{author}{\bibfnamefont{J.~D.~E.} \bibnamefont{Creighton}},
  \bibnamefont{and} \bibinfo{author}{\bibfnamefont{A.~G.}
  \bibnamefont{Wiseman}}, \bibinfo{journal}{Class. Quant. Grav.}
  \textbf{\bibinfo{volume}{21}}, \bibinfo{pages}{S1775} (\bibinfo{year}{2004}).

\bibitem[{\citenamefont{Kalogera et~al.}(2001)\citenamefont{Kalogera, Narayan,
  Spergel, and Taylor}}]{Kalogera:2000dz}
\bibinfo{author}{\bibfnamefont{V.}~\bibnamefont{Kalogera}},
  \bibinfo{author}{\bibfnamefont{R.}~\bibnamefont{Narayan}},
  \bibinfo{author}{\bibfnamefont{D.~N.} \bibnamefont{Spergel}},
  \bibnamefont{and} \bibinfo{author}{\bibfnamefont{J.~H.}
  \bibnamefont{Taylor}}, \bibinfo{journal}{Astrophys. J.}
  \textbf{\bibinfo{volume}{556}}, \bibinfo{pages}{340} (\bibinfo{year}{2001}).

\bibitem[{\citenamefont{Thorsett and Chakrabarty}(1999)}]{Thorsett:1998uc}
\bibinfo{author}{\bibfnamefont{S.~E.} \bibnamefont{Thorsett}} \bibnamefont{and}
  \bibinfo{author}{\bibfnamefont{D.}~\bibnamefont{Chakrabarty}},
  \bibinfo{journal}{Astrophys. J.} \textbf{\bibinfo{volume}{512}},
  \bibinfo{pages}{288} (\bibinfo{year}{1999}).

\bibitem[{\citenamefont{Adhikari et~al.}(2003)\citenamefont{Adhikari,
  Gonz{\'a}lez, Landry, and O'Reilly}}]{Gonzalez:2002}
\bibinfo{author}{\bibfnamefont{R.}~\bibnamefont{Adhikari}},
  \bibinfo{author}{\bibfnamefont{G.}~\bibnamefont{Gonz{\'a}lez}},
  \bibinfo{author}{\bibfnamefont{M.}~\bibnamefont{Landry}}, \bibnamefont{and}
  \bibinfo{author}{\bibfnamefont{B.}~\bibnamefont{O'Reilly}},
  \bibinfo{journal}{Class. Quant. Grav.} \textbf{\bibinfo{volume}{20}},
  \bibinfo{pages}{S903} (\bibinfo{year}{2003}).

\bibitem[{\citenamefont{Brown (for~the LIGO
  Scientific~Collaboration)}(2004)}]{Brown:2004}
\bibinfo{author}{\bibfnamefont{D.~A.} \bibnamefont{Brown (for~the LIGO
  Scientific~Collaboration)}}, \bibinfo{journal}{Class. Quant. Grav.}
  \textbf{\bibinfo{volume}{21}}, \bibinfo{pages}{S797} (\bibinfo{year}{2004}).

\bibitem[{\citenamefont{Gonz{\'a}lez et~al.}(2004)\citenamefont{Gonz{\'a}lez,
  Landry, and O'Reilly}}]{Gonzalez:2004}
\bibinfo{author}{\bibfnamefont{G.}~\bibnamefont{Gonz{\'a}lez}},
  \bibinfo{author}{\bibfnamefont{M.}~\bibnamefont{Landry}}, \bibnamefont{and}
  \bibinfo{author}{\bibfnamefont{B.}~\bibnamefont{O'Reilly}},
  \bibinfo{type}{Tech. Rep.} \bibinfo{number}{LIGO-T040060-00-D},
  \bibinfo{institution}{LIGO Project} (\bibinfo{year}{2004}),
  \urlprefix\url{http://www.ligo.caltech.edu/docs/T/T040060-00.pdf}.

\bibitem[{\citenamefont{Abbott et~al.}(2005{\natexlab{c}})}]{ligotama}
\bibinfo{author}{\bibfnamefont{B.}~\bibnamefont{Abbott}} \bibnamefont{et~al.}
  (\bibinfo{collaboration}{LIGO-TAMA}), \bibinfo{journal}{in preparation}
  (\bibinfo{year}{2005}{\natexlab{c}}).

\end{thebibliography}
\end{document}